\documentclass[twocolumn]{aastex63}
\usepackage{CJK}

\newcommand{\bjdtdb}{\ensuremath{\rm {BJD_{TDB}}}}

\newcommand{\fave}{\langle F \rangle}
\newcommand{\fluxcgs}{10$^9$ erg s$^{-1}$ cm$^{-2}$}

\newcommand{\cmfrho}{{$\rm CMF_{\rho}$}}
\newcommand{\cmfstar}{{$\rm CMF_{\star}$}}

\newcommand{\vsini}{\ensuremath{v\sin{i_*}}}

\newcommand{\dens}{g cm$^{-3}$}

\def\wotan{{\tt W\={o}tan}}
\def\k2{{\it K2}}
\shorttitle{A re-analysis of K2-106b}
\shortauthors{Rodr\'iguez Mart\'inez et al.}

\begin{document}
\begin{CJK*}{UTF8}{gbsn}

\title{A Reanalysis of the Composition of K2-106b: an Ultra-short Period Super-Mercury Candidate}

\correspondingauthor{Romy Rodr\'iguez Mart\'inez}
\email{rodriguezmartinez.2@osu.edu}

\author[0000-0003-1445-9923]{Romy Rodr\'iguez Mart\'inez}
\affiliation{Department of Astronomy, The Ohio State University, 140 W. 18th Avenue, Columbus, OH 43210, USA}

\author[0000-0003-0395-9869]{B. Scott Gaudi}
\affiliation{Department of Astronomy, The Ohio State University, 140 W. 18th Avenue, Columbus, OH 43210, USA}

\author[0000-0003-3570-422X]{
Joseph G. Schulze}
\affiliation{School of Earth Sciences, The Ohio State University, 125 South Oval Mall, Columbus OH, 43210, USA}
\author[0000-0002-9147-7925]{Lorena Acu\~na}
\affiliation{Aix-Marseille Univ., CNRS, CNES, LAM, Marseille, France}

\author{Jared Kolecki}
\affiliation{Department of Astronomy, The Ohio State University, 140 W. 18th Avenue, Columbus, OH 43210, USA}

\author{Jennifer A. Johnson}
\affiliation{Department of Astronomy, The Ohio State University, 140 W. 18th Avenue, Columbus, OH 43210, USA}

\author[0000-0002-8823-8237
]{Anusha Pai Asnodkar}
\affiliation{Department of Astronomy, The Ohio State University, 140 W. 18th Avenue, Columbus, OH 43210, USA}

\author[0000-0001-8153-639X]{Kiersten M. Boley}
\altaffiliation{NSF Graduate Research Fellow}
\affiliation{Department of Astronomy, The Ohio State University, 140 W. 18th Avenue, Columbus, OH 43210, USA}

\author{Magali Deleuil}
\affiliation{Aix-Marseille Univ., CNRS, CNES, LAM, Marseille, France}

\author{Olivier Mousis}
\affiliation{Aix-Marseille Univ., CNRS, CNES, LAM, Marseille, France}

\author[0000-0001-5753-2532]{Wendy R. Panero}
\affiliation{School of Earth Sciences, The Ohio State University, 125 South Oval Mall, Columbus OH, 43210, USA}

\author[0000-0002-4361-8885]{Ji Wang (王吉)}
\affiliation{Department of Astronomy, The Ohio State University, 140 W. 18th Avenue, Columbus, OH 43210, USA}

\begin{abstract}

We present a reanalysis of the K2-106 transiting planetary system, with a focus on the composition of K2-106b, an ultra-short period, super-Mercury candidate. We globally model existing photometric and radial velocity data and derive a planetary mass and radius for K2-106b of $M_{p} = 8.53\pm1.02~M_{\oplus}$ and $R_{p} = 1.71^{+0.069}_{-0.057}~R_{\oplus}$, which leads to a density of $\rho_{p} = 9.4^{+1.6}_{-1.5}$~\dens, a significantly lower value than previously reported in the literature. We use planet interior models that assume a two-layer planet comprised of a liquid, pure Fe core and iron-free, $\rm MgSiO_{3}$ mantle, and we determine the range of core mass fractions that are consistent with the observed mass and radius. We use existing high-resolution spectra of the host star to derive Fe/Mg/Si abundances ([Fe/H]$=-0.03 \pm 0.01$, [Mg/H]$= 0.04 \pm 0.02$, [Si/H]$=0.03 \pm 0.06$) to infer the composition of K2-106b. We find that although K2-106b has a high density and core mass fraction ($44^{+12}_{-15}\%$) compared to the Earth (33\%), its composition is consistent with what is expected assuming that it reflects the relative refractory abundances of its host star. K2-106b is therefore unlikely to be a super-Mercury, as has been suggested in previous literature.

\end{abstract}

\keywords{Exoplanet systems, planetary interior -- techniques: transit photometry -- techniques: radial velocity -- exoplanets: rocky planets} 

\section{Introduction} \label{sec:intro}

Ultra-short period planets
(USPs) are a class of planets characterized by orbital periods shorter than one day \citep{Winn:2018}. These planets are typically smaller than $2 R_{\oplus}$ and have relatively high densities consistent with terrestrial compositions and thin (or no) atmospheres. Some well-known USPs include CoRoT-7b \citep{Leger:2009}, Kepler-10b \citep{Batalha:2010}, 55 Cancri e \citep{Ligi:2016}, and Kepler-78b \citep{Howard:2013, Pepe:2013}. NASA's Transiting Exoplanet Survey Satellite (TESS; \citealt{Ricker:2015}) has already detected dozens of USPs, and it is expected to contribute many more.

USPs have occurrence rates similar to hot Jupiters, of approximately 0.5\%, 0.8\%, and 1.1\% around G, K, and M dwarfs, respectively \citep{Sanchis-Ojeda:2014}.  And similarly to hot Jupiters, their origin and formation is still the subject of debate. One early formation theory proposed that USPs were originally hot Jupiters that were stripped of their atmospheres by tidal disruption and stellar irradiation and thus are the remnant cores of such planets. However, this theory is inconsistent with the observation that the host stars of USPs and hot Jupiters have very different metallicity distributions, suggesting that they are different underlying populations \citep{Winn:2017}. Instead, USPs are more likely to be remnants of lower-mass, Neptune-sized gas dwarfs that lost their primordial H/He atmospheres due to low atmospheric energy or surface gravity and photoevaporation \citep{Owen:2017}. It is also possible that these planets formed farther out in a region of more solids in the protoplanetary disk and then migrated to their observed present locations (see \citealt{Dai:2019} and references therein). The characterization of USPs provides valuable insights into planet formation and present a unique laboratory to investigate the exposed, remnant cores of gas dwarfs.

Small exoplanets ($R_{p} \lesssim 3R_{\oplus}$ and $M_{p} \lesssim 10 M_{\oplus}$) show a rich diversity of densities and compositions (see, e.g., \citealt{Jontof-Hutter:2019}). While many of these appear to follow Earth-like compositions (comprised primarily of $\sim$20\% iron and $\sim$80\% rock; e.g., \citealt{Dressing:2015}), a few more exotic classes of exoplanets have begun to emerge. One of these are super-Mercuries: typically dense planets that appear to have interior compositions consistent with Mercury's. While Venus, Earth and Mars have a core mass fraction (CMF) of approximately 30\%, Mercury has a much larger CMF of $\sim$70\% \citep{Smith:2012} that is inconsistent with what we expect based on the relative refractory (or rock-forming) elemental abundances of the Sun. Here, we explicitly define a super-Mercury as a planet that is iron enriched relative to the Mg and Si of the protoplanetary disk from which it formed, regardless of its bulk density. Several exoplanets have been identified as super-Mercuries so far, including K2-229b \citep{Santerne:2018}, Kepler-107c \citep{Bonomo:2019,Schulze:2021}, GJ 367b \citep{Lam:2021}, HD 137496b \citep{Silva:2022}, and lastly K2-106b \citep{Adams:2017}, which is the subject of this paper.

The origin of super-Mercuries (and indeed, Mercury) remains largely a mystery, but several hypotheses have been proposed to explain their formation. One of these is that the inner regions of protoplanetary disks are iron enriched relative to silicates, facilitating the formation of highly dense planets close to the star \citep{Lewis:1972,Aguichine:2020,Scora:2020,Adibekyan:2021}. One example of such a compositional sorting mechanism is photophoresis, a well-known aerosol physics force only recently considered in planet formation contexts (e.g., \citealt{Wurm:2013}). Outward radial acceleration due to photophoresis is larger for lower-conductivity/density materials (such as silicates) relative to their higher-conductivity/density counterparts (metals), resulting in metal-silicate gradients with metal-rich regions close to the star. Other theories invoke either small or giant late-stage impacts that remove part of the planet's mantle, thus leaving behind a relatively larger iron core \citep{Leinhardt:2012,Marcus:2010}, and mantle evaporation \citep{Cameron:1985}. The detection and characterization of Mercury-like planets will improve our understanding of the formation and evolutionary pathways that lead to the formation of such objects. More generally, it will illuminate the compositional diversity of low-mass planets and the chemical links between rocky planets and their host stars (e.g., \citealt{Dorn:2015, unterborn:2016, Brugger:2017, Schulze:2021, Plotnykov:2020, Liu:2020, Adibekyan:2021}).

In this paper, we focus on K2-106 (EPIC 220674823, TIC 266015990), a G-type star with $V = 12.10$ mag located at RA 00$^{h}$52$^{m}$19$\fs$14, Dec $+$10$\degr$47$\arcmin$40$\arcsec$90, at a distance of 244.7$\pm$ 0.001 pc (see Table~\ref{tbl:LitProps}). \citet{Adams:2017} identified two planets in the system in \k2 data: a USP at a period of 0.57 days (K2-106b) and a more massive, outer companion with $P = 13.3$ days (K2-106c). They statistically validated planets b and c and placed upper limits on their masses of $M_{b} < 0.43 M_{\rm Jup}$ and $M_{c} < 1.22 M_{\rm Jup}$. \citet{Sinukoff:2017} revisited the system and confirmed the masses of the planets with radial velocities by obtaining high-resolution spectra from the HIRES spectrograph, deriving a mass and radius for K2-106b of $M_{p} = 9.0 \pm 1.6 M_{\oplus}$ and $R_p = 1.82^{+0.20}_{-0.14} R_{\oplus}$, leading to a bulk density of $\rho_{p} = 8.57^{+4.64}_{-2.80}$ \dens. \citet{Guenther:2017} (hereafter G17) reanalyzed this system and improved upon previous mass estimates by collecting extensive radial velocities. They determined a planetary mass and radius of $M_{p} = 8.36 \pm 0.95 M_{\oplus}$ and $R_p = 1.52 \pm 0.16 R_{\oplus}$, implying a bulk density of $\rho_{p} = 13.1 ^{+5.4}_{-3.6}$ \dens. G17 also derived a CMF of $80^{+20}_{-30}$\% and therefore concluded that K2-106b is a super-Mercury. Motivated by the improved precision of the parallaxes from the second {\it Gaia} data release \citep{Gaia:2018}, \citet{Dai:2019} reanalyzed a sample of 11 USPs, including K2-106b, with the goal of improving their stellar and planetary properties. \citet{Dai:2019} derived $M_{p} = 7.72^{+0.80}_{-0.79} M_{\oplus}$ and $R_p = 1.712 \pm 0.068 R_{\oplus}$, inferring a bulk density of $\rho_{p} = 8.5 \pm 1.9$ \dens. They also derived a lower CMF than G17, which led them to conclude that K2-106b is likely not as iron-enriched as previously thought.

To accurately and precisely determine the properties and structure of small exoplanets, we need accurate and precise planetary masses and radii, which in turn critically depend on both the precision in the stellar parameters and the methods used to derive them. Traditionally, measurements of stellar masses and radii have relied upon theoretical stellar evolutionary tracks and/or empirical relationships between stellar mass and radius. However, these methods typically make assumptions that are not necessarily valid for the types of stars under study, and there are errors arising from uncertainties in the physics of stellar structure and evolution. Moreover, studies have demonstrated that the uncertainty in stellar parameters derived using constraints from stellar evolutionary models (such as the  MESA Isochrones and Stellar Tracks (MIST; \citealt{Dotter:2016,Choi:2016,Paxton:2011,Paxton:2013,Paxton:2015}), the Yonsei Yale (YY) evolutionary models \citep{Yi:2001} or the \citet{Torres:2010} empirical relationships) are typically underestimated \citep{Tayar:2020} and therefore propagate to underestimated planet property uncertainties. In addition, these different stellar models give inconsistent answers (\citealt{Tayar:2020}, Duck et al. in preparation); hence the importance of deriving model-independent stellar and planetary parameters (see \citet{Stassun:2017} for a discussion of how to empirically determine stellar parameters for the host stars of transiting planets).

Using the analytic approximations for the uncertainty in various physical properties of transiting exoplanets from \citet{RodriguezMartinez:2021}, we performed a preliminary analysis of the properties of the host K2-106. We found that the uncertainty in the density of K2-106b could be reduced by a factor of $\sim$2 compared to the value found by G17, provided that we achieved a comparable precision to G17 in the transit and RV observables $T$ (transit duration), $P$ (period), $K_{\star}$ (RV semiamplitude), $\delta$ (transit depth), and $\tau$ (ingress/egress time). In this paper, we expand upon the work of G17, \citet{Dai:2019}, and \citet{RodriguezMartinez:2021}, and we revisit K2-106 using constraints from stellar evolutionary models as priors, and without, in both cases using a more precise parallax from {\it Gaia} DR3 \citep{GaiaDR3} (which is a factor of $\sim$3 more precise than the one from DR2) to improve upon the mass, radius, and density of K2-106b and thus better constrain the composition of this planet. Being a super-Mercury candidate, a USP planet, and in a multi-planet system, K2-106b is an excellent target to study the formation and evolution of these rare objects. This system is also particularly interesting because it has one planet that is highly irradiated and another small planet farther out, making it a good target to study the evolution of atmospheres and atmospheric escape, as well as the compositional diversity of planets in the same system.

\section{Data} \label{sec:data}
\subsection{Photometry}

\begin{figure}[!ht]
\vspace{.1in}
\centering
\includegraphics[width=1\linewidth]{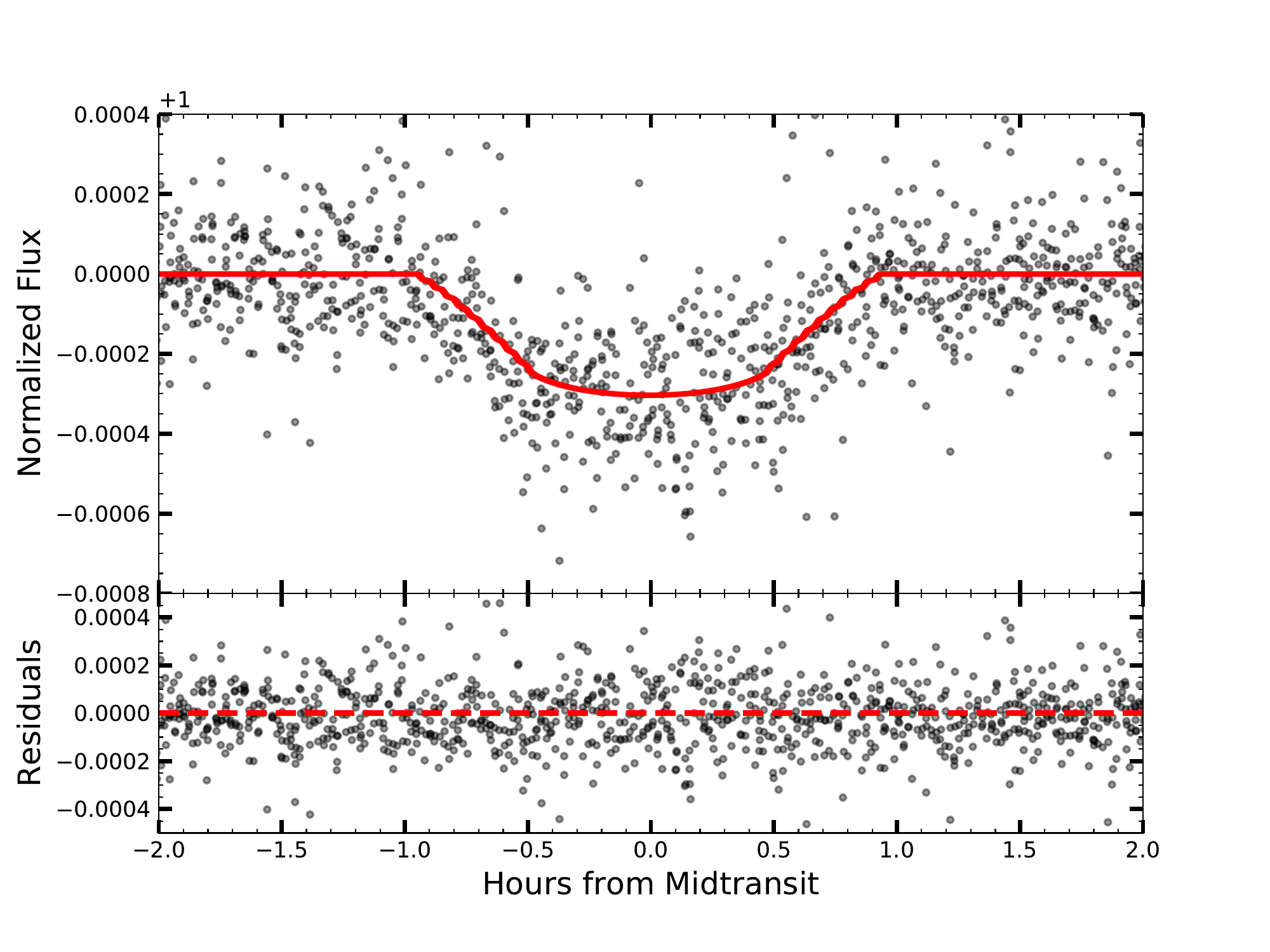}
\includegraphics[width=1\linewidth]{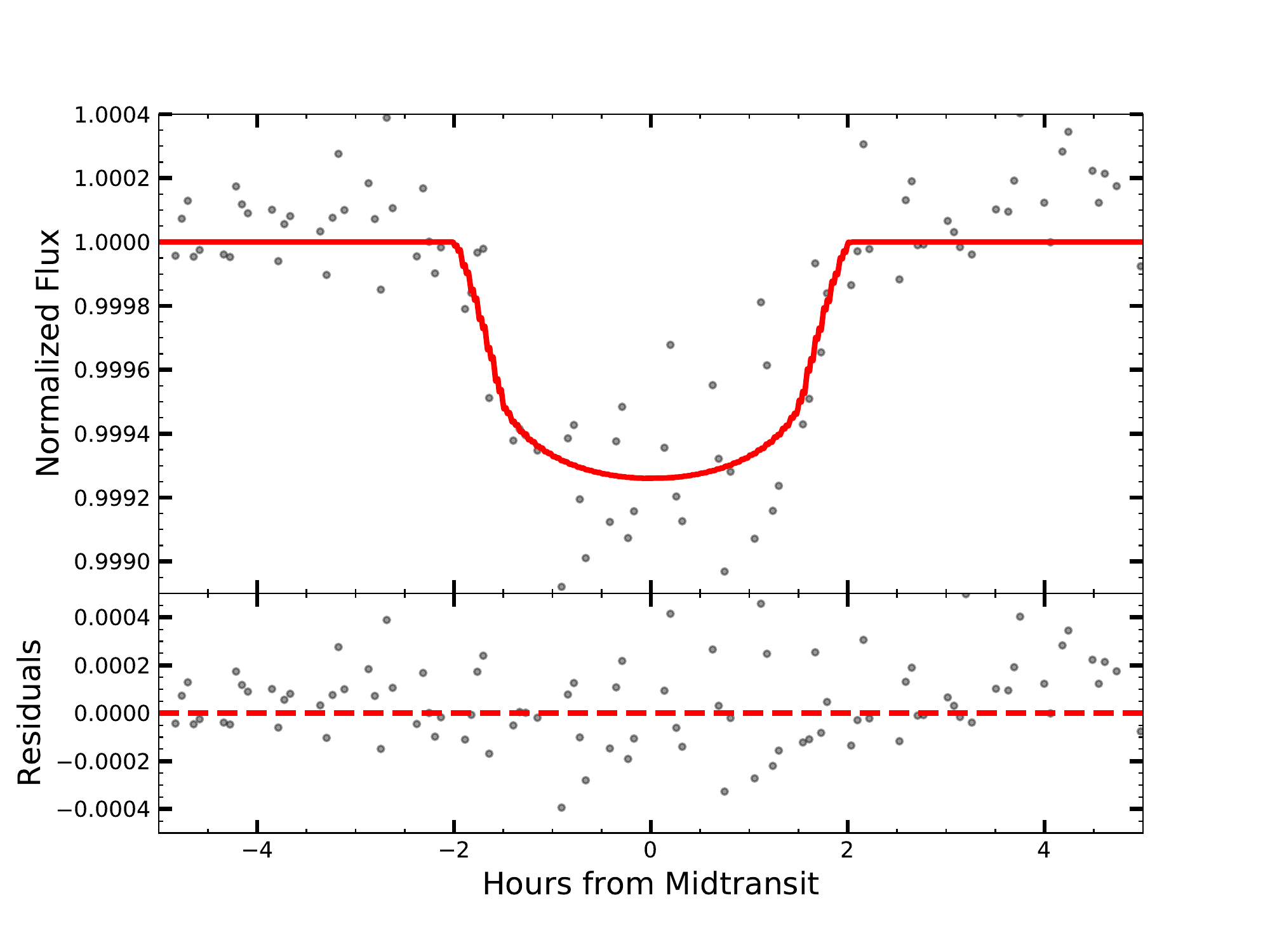}
\caption{Phase-folded $K2$ transit light curves of K2-106b \textbf{(top)} and K2-106c \textbf{(bottom)}. The dark points show the measurements and the red lines show the best-fit model, with residuals plotted below the light curves.}
\label{fig:lightcurves} 
\end{figure}

To characterize the K2-106 system, we used existing photometry from the NASA \k2 mission \citep{howell:2014}. K2-106 was observed in campaign 8 of \k2. We retrieved the 30-minute cadence, \k2 light curve from the K2 Extracted Light Curves (K2SFF) database \citep{Vanderburg:2014}, which has already been corrected for instrumental systematics. The light curve was detrended and normalized with the \wotan~package \citep{Hippke:2014} using the Tukey's biweight method. The phase-folded transit light curves for K2-106b and K2-106c are shown in Figure~\ref{fig:lightcurves}.

K2-106 was observed by TESS between August 20 and September 16 2021, in Sector 42 by camera 3 in cycle 4. The TESS data was processed through the TESS Science Processing Operations Center (SPOC) pipeline, and the light curve was retrieved via the Barbara A. Mikulski Archive for Space Telescopes (MAST). We removed any data points that were flagged for quality issues. We detrended the light curve using \wotan~before conducting a transit search using Transit Least Squares. The search produced a planet detection with a signal-to-noise ratio SNR $\sim$3. Thus, the TESS light curve for K2-106 does not provide  useful additional constraints and therefore we did not use it in this work.

\begin{table}
\scriptsize
\setlength{\tabcolsep}{2pt}
\centering
\caption{Literature Properties for K2-106}
\begin{tabular}{llcc}
  \hline
  \hline
\multicolumn{3}{c}{EPIC 220674823,} \\
\multicolumn{3}{c}{TIC 266015990, TYC 608-458-1} \\
\hline
\hline
Parameter & Description & Value & Source\\
\hline 
$\alpha_{J2000}$\dotfill	&Right Ascension (RA)\dotfill & 00:52:19.14& 1	\\
$\delta_{J2000}$\dotfill	&Declination (Dec)\dotfill & +10:47:40.90& 1	\\
$l$\dotfill     & Galactic Longitude\dotfill & 	123.2839$^\circ$ & 1\\
$b$\dotfill     & Galactic Latitude\dotfill & -52.0764$^\circ$ & 1\\
\\
${\rm G}$\dotfill     & Gaia $G$ mag.\dotfill     &11.953 $\pm$ 0.02& 2\\
${\rm G_{BP}}$\dotfill     & Gaia $G$ mag.\dotfill     &12.355 $\pm$ 0.02& 2\\
${\rm G_{RP}}$\dotfill     & Gaia $G$ mag.\dotfill     &11.403 $\pm$ 0.02& 2\\
\\
J\dotfill			& 2MASS J mag.\dotfill & 10.770 $\pm$ 0.023	& 3	\\
H\dotfill			& 2MASS H mag.\dotfill & 10.454 $\pm$ 0.026	    &  3	\\
K$_S$\dotfill			& 2MASS ${\rm K_S}$ mag.\dotfill & 10.344 $\pm$  0.021&  3	\\
\\
\textit{WISE 1}\dotfill		& \textit{WISE 1} mag.\dotfill & 10.299 $\pm$ 0.030 & 4	\\
\textit{WISE 2}\dotfill		& \textit{WISE 2} mag.\dotfill & 10.355 $\pm$ 0.030 &  4 \\
\textit{WISE 3}\dotfill		& \textit{WISE 3} mag.\dotfill &  10.380 $\pm$ 0.091& 4	\\

$\pi^\dagger$\dotfill & Gaia Parallax (mas) \dotfill & 4.121 $\pm$  0.049 &  2 \\
$d$\dotfill & Distance (pc)\dotfill & 244.7 $\pm$ 0.001 & 2 \\
 \\
\hline
\end{tabular}
\begin{flushleft}
 \footnotesize{ \textbf{\textsc{NOTES:}}
RA and Dec are in epoch J2000. The coordinates come from Vizier. $\dagger$The parallax has been corrected for the -0.036 $\mu$as offset as reported by \citet{Lindegren:2018}.\\
References are: $^1$Vizier, $^2$\citet{GaiaDR3}, $^3$\citet{Cutri:2003}, $^4$\citet{Zacharias:2004}.\\
}
\end{flushleft}
\label{tbl:LitProps}
\end{table}

\subsection{Radial Velocities}

G17 determined that K2-106 is inactive based on their derivation of an average chromospheric activity index of log $R'_{\rm HK} = -5.04 \pm 0.19$, which is near the mininum value of the chromospheric activity index for stars with solar metallicity ($\sim$ log $R'_{\rm HK} = -5.08$). They also and derive a slow rotation for the star of $\vsini = 2.8 \pm 0.35$ km/s, thus concluding that the star is inactive.

To constrain the masses of K2-106b and K2-106c, we used multiple radial velocity data sets collected by G17 as well as the HIRES RV data from \citep{Sinukoff:2017}. We briefly describe the instruments and the data below.

\textbf{PFS}: We used 13 spectra of K2-106 from the Carnegie Planet Finder Spectrograph  (PFS; \citealt{, Crane:2006,Crane:2010}) on the 6.5 m Magellan/Clay Telescope, taken between August 14, 2016 and January 14, 2017. PFS has a resolving power of $\lambda/\Delta \lambda \sim 76,000$. The exposures ranged from 20--40 min and had an SNR of 50--140 per pixel. The relative velocities were extracted from the spectrum using the techniques in \citet{Butler:1996}. 

\textbf{HDS}: We used 3 RV measurements from the High Dispersion Spectrograph (HDS; \citealt{Noguchi:2002}) located on the 8.2 m Subaru Telescope, obtained between October 12--14, 2016. The telescope has a resolving power of $\lambda/\Delta \lambda \sim 85,000$ and a typical SNR of 70--80 per pixel. G17 notes that, as with PFS, the HDS RVs were measured relative to a template spectrum taken by the same instrument without the iodine cell. 

\begin{figure}[!ht]
\vspace{.1in}
\centering
\includegraphics[width=1\linewidth, trim={2.5cm 13cm 8.5cm 8cm}]{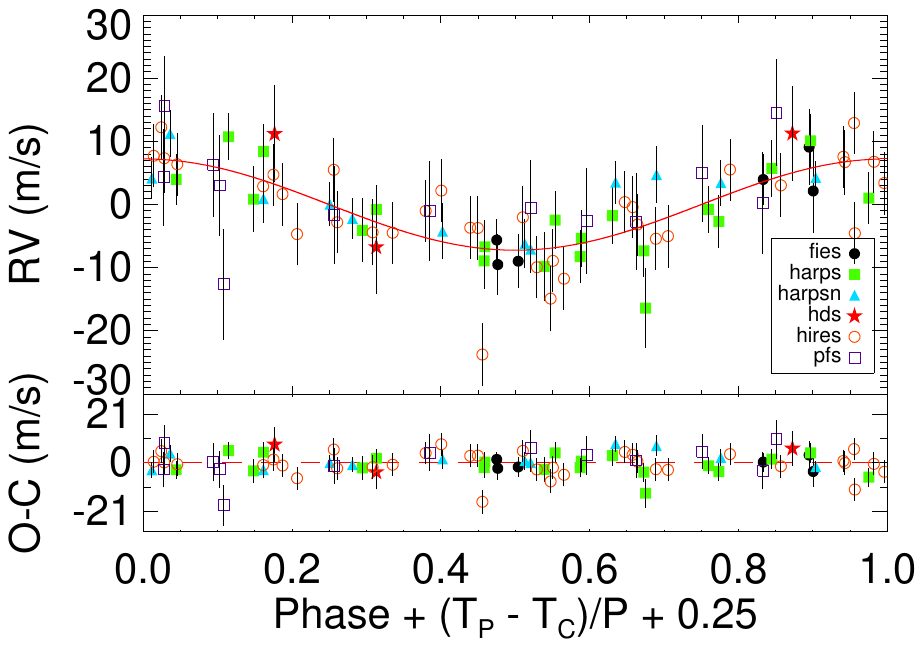}
\includegraphics[width=1\linewidth, trim={2.5cm 13cm 8.5cm 8cm}]{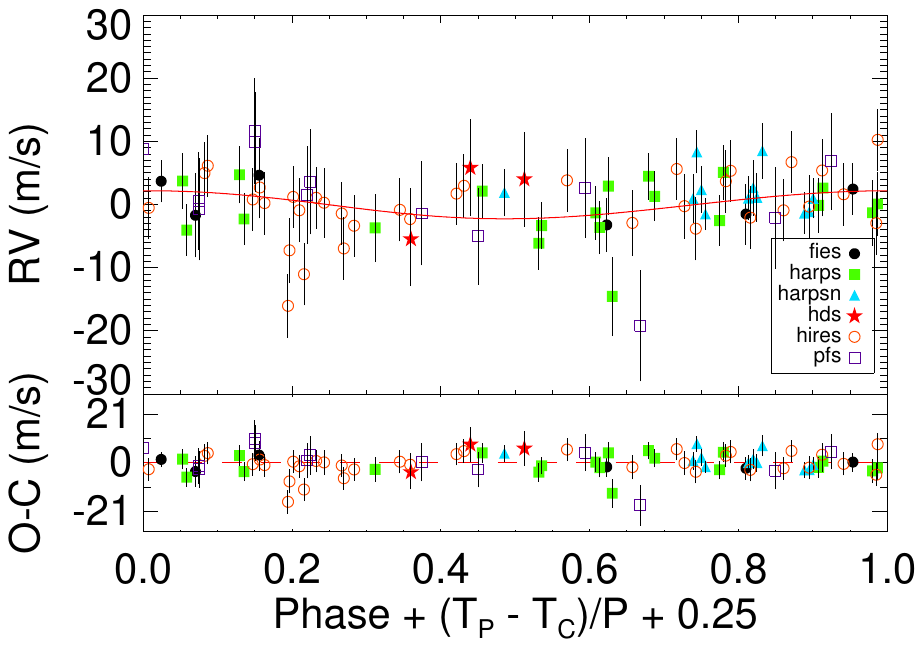}
\caption{Phase-folded radial velocity measurements of K2-106b \textbf{(top)} and K2-106c \textbf{(bottom)}. The red line shows the best-fit {\tt EXOFASTv2} model, with residuals plotted below the RV curves. $\rm T_P$ is the time of periastron, while $\rm T_C$ is the time of conjuction or transit.}
\label{fig:RVs} 
\end{figure}

\textbf{FIES}: We used 6 RV measurements from the FIbre-fed Echelle Spectrograph (FIES; \citealt{Frandsen:1999,Telting:2014}) on the 2.56 m Nordic Optical Telescope (NOT) at the Observatorio del Roque de los Muchachos, in La Palma, Spain. The observations were taken from October 5 to November 25, 2016. They used the 1.3" high-resolution fiber, with $\lambda/\Delta \lambda \sim 67,000$ and they reduced the spectra using standard IRAF and IDL routines.

\textbf{HIRES}: We included 35 RV measurements taken by \citet{Sinukoff:2017} from the High Resolution Echelle Spectrometer (HIRES) instrument at Keck \citep{Vogt:1994}, which has a resolving power of $\sim$100,000. 

\subsection{HARPS and HARPS-N data}

We used 20 RV measurements from the High Accuracy Radial velocity Planet Searcher (HARPS; \citealt{Mayor:2003}) spectrograph on the 3.6 m ESO Telescope at La Silla, Chile. Additionally, we used 12 RV measurements from the HARPS-N instrument at the 3.58-m Telescopio Nazionale \textit{Galileo} (TNG) at La Palma \citep{Cosentino:2012}. The HARPS spectra were obtained from October 25, 2016 to November 27, 2016 and the HARPS-N data between October 30, 2016 to January 28, 2017. Both instruments have a resolving power of $\lambda/\Delta \lambda \sim 115,000$. The spectra were then reduced with the dedicated HARPS and HARPS-N pipelines. 

In total, we used 89 RV measurements for our analysis. We refer the reader to G17 and \citet{Sinukoff:2017} for a more detailed description of all the RV measurements and their reduction. The phase-folded RV curves of the system are plotted in Figure~\ref{fig:RVs}.

\begin{table*}
\scriptsize
\centering
\caption{Median values and 68\% confidence interval for the physical parameters of K2-106b and K2-106c from the global fit using constraints from MIST. \label{tab:mist_table}}
\begin{tabular}{llcccc}
  \hline
  \hline
Parameter & Description (Units) & Values & Values & & \\
&& K2-106b & K2-106c \\
\hline
~~~~$P$\dotfill &Period (days)\dotfill &$0.571302^{+0.000015}_{-0.000016}$&$13.3393^{+0.0014}_{-0.0015}$\\
~~~~$R_P$\dotfill &Radius ($R_{\oplus}$)\dotfill &$1.710^{+0.069}_{-0.057}$&$2.726^{+0.134}_{-0.1234}$\\
~~~~$M_P$\dotfill &Mass ($M_{\oplus}$)\dotfill &$8.53\pm1.02$&$
5.919^{+3.50}_{-3.08}$\\
~~~~$T_C$\dotfill &Time of conjunction (\bjdtdb)\dotfill & $2457394.0106\pm0.0013$ & $2457405.7345^{+0.0039}_{-0.0037}$\\
~~~~$a$\dotfill &Semi-major axis (AU)\dotfill &$0.01332^{+0.00028}_{-0.00029}$&$0.1088^{+0.0023}_{-0.0024}$\\
~~~~$i$\dotfill &Inclination (Degrees)\dotfill &$84.2^{+3.5}_{-2.9}$ & $88.74^{+0.60}_{-0.24}$\\
~~~~$T_{eq}$\dotfill &Equilibrium temperature (K)\dotfill &$2275^{+36}_{-32}$&$796^{+13}_{-11}$\\
~~~~$K_{\star}$\dotfill &RV semi-amplitude (m/s)\dotfill &$6.7^{+0.75}_{-0.76}$&$1.66^{+0.94}_{-0.86}$\\
~~~~$R_P/R_*$\dotfill &Radius of planet in stellar radii \dotfill &$0.01604^{+0.00043}_{-0.00040}$&$0.02555^{+0.001}_{-0.00092}$\\
~~~~$a/R_*$\dotfill &Semi-major axis in stellar radii \dotfill &$2.931^{+0.089}_{-0.098}$&$23.95^{+0.73}_{-0.8}$\\
~~~~$\delta$\dotfill &Transit depth (fraction)\dotfill &$0.000257^{+0.000014}_{-0.000013}$&$0.000653^{+0.000055}_{-0.000046}$\\
~~~~$\tau$\dotfill &Ingress/egress transit duration (days)\dotfill &$0.001103^{+0.00013}_{-0.000077}$&$0.005^{+0.0014}_{-0.0011}$\\

~~~~$e$\dotfill & Eccentricity \dotfill & 0 (fixed) & $ 0.13^{+0.12}_{-0.08}$\\

~~~~$T_{14}$\dotfill &Total transit duration (days)\dotfill &$0.0615^{+0.0018}_{-0.0019}$&$0.1498\pm0.0047$\\
~~~~$T_{FWHM}$\dotfill &FWHM transit duration (days)\dotfill &$0.0604^{+0.0018}_{-0.002}$&$0.1447^{+0.0048}_{-0.0047}$\\
~~~~$b$\dotfill &Transit Impact parameter \dotfill &$0.3^{+0.13}_{-0.18}$&$0.5^{+0.14}_{-0.27}$\\
~~~~$\rho_P$\dotfill &Density (cgs)\dotfill &$9.4^{+1.6}_{-1.5}$&$1.58^{+0.96}_{-0.84}$\\
~~~~$logg_P$\dotfill &Surface gravity \dotfill &$3.455^{+0.059}_{-0.065}$&$2.89^{+0.2}_{-0.32}$\\
~~~~$\Theta$\dotfill &Safronov Number \dotfill &$0.00484
\pm0.00057$&$0.0172^{+0.0098}_{-0.009}$\\
~~~~$\fave$\dotfill &Incident Flux (\fluxcgs)\dotfill &$6.08^{+0.39}_{-0.33}$&$0.0886^{+0.006}_{-0.0055}$\\
~~~~$T_S$\dotfill &Time of eclipse (\bjdtdb)\dotfill &$2457394.2962\pm0.0013$ &$2457412.17^{+0.76}_{-1.1}$\\
~~~~$T_A$\dotfill &Time of Ascending Node (\bjdtdb)\dotfill &$2457394.4391\pm0.0013$&$2457402.44^{+0.62}_{-0.61}$\\
~~~~$T_D$\dotfill &Time of Descending Node (\bjdtdb)\dotfill &$2457394.1534\pm0.0013$&$2457408.75^{+0.51}_{-0.70}$\\
~~~~$M_P\sin i$\dotfill &Minimum mass ($M_{\oplus}$)\dotfill &$8.497\pm1.01$&$5.919^{+3.50}_{-3.08}$\\
~~~~$M_P/M_*$\dotfill &Mass ratio \dotfill &$0.0000266\pm0.0000031$&$0.0000184^{+0.00001}_{-0.0000096}$\\
~~~~$d/R_*$\dotfill &Separation at mid transit \dotfill &$2.931^{+0.089}_{-0.098}$&$22.5^{+2.4}_{-2.9}$\\
~~~~$P_T$\dotfill &A priori non-grazing transit prob \dotfill &$0.3357^{+0.012}_{-0.0099}$&$0.0433^{+0.0065}_{-0.0042}$\\
~~~~$P_{T,G}$\dotfill &A priori transit prob \dotfill &$0.347^{+0.012}_{-0.01}$&$0.0455^{+0.0067}_{-0.0044}$\\
\hline
\end{tabular}
\label{tab:exofast_planetary}
\end{table*}

\begin{table*}
\scriptsize
\centering
\caption{Median values and 68\% confidence interval for the physical parameters of K2-106b and K2-106c from the model-independent global fit. \label{tab:nomodel_table}}
\begin{tabular}{llcccc}
  \hline
  \hline
Parameter & Description (Units) & Values & Values & & \\
&& K2-106b & K2-106c \\
\hline
~~~~$P$\dotfill &Period (days)\dotfill &$0.571301^{+0.000015}_{-0.000016}$&$13.3393\pm0.0015$\\
~~~~$R_P$\dotfill &Radius ($R_{\oplus}$)\dotfill &$1.75^{+0.086}_{-0.068}$&$2.782^{+0.157}_{-0.134}$\\
~~~~$M_P$\dotfill &Mass ($M_{\oplus}$)\dotfill &$8.69^{+1.69}_{-1.90}$&$
6.046^{+3.81}_{-3.18}$\\
~~~~$T_C$\dotfill &Time of conjunction (\bjdtdb)\dotfill & $2457394.0106\pm0.0013$ & $2457405.7346^{+0.0039}_{-0.0037}$\\
~~~~$a$\dotfill &Semi-major axis (AU)\dotfill &$0.01357^{+0.0009}_{-0.0016}$&$0.1108^{+0.0074}_{-0.013}$\\
~~~~$i$\dotfill &Inclination (Degrees)\dotfill &$84.4^{+4}_{-6.8}$ & $88.76^{+0.61}_{-0.53}$\\
~~~~$e$\dotfill &eccentricity (Degrees)\dotfill &0 (fixed) & $0.131^{+0.12}_{-0.089}$\\
~~~~$T_{eq}$\dotfill &Equilibrium temperature (K)\dotfill &$2261^{+140}_{-70}$&$791^{+50}_{-25}$\\
~~~~$K_{\star}$\dotfill &RV semi-amplitude (m/s)\dotfill &$6.68 \pm 0.75$&$1.69^{+0.93}_{-0.88}$\\
~~~~$R_P/R_*$\dotfill &Radius of planet in stellar radii \dotfill &$0.01617^{+0.0006}_{-0.00043}$&$0.02565^{+0.0013}_{-0.00098}$\\
~~~~$a/R_*$\dotfill &Semi-major axis in stellar radii \dotfill &$2.95^{+0.16}_{-0.34}$&$24.1^{+1.3}_{-2.8}$\\
~~~~$\delta$\dotfill &Transit depth (fraction)\dotfill &$0.000261^{+0.00002}_{-0.000014}$&$0.000658^{+0.000066}_{-0.000049}$\\
~~~~$\tau$\dotfill &Ingress/egress transit duration (days)\dotfill &$0.001091^{+0.00042}_{-0.000098}$&$0.005^{+0.0022}_{-0.0012}$\\
~~~~$T_{14}$\dotfill &Total transit duration (days)\dotfill &$0.0609^{+0.0022}_{-0.0021}$&$0.1498\pm0.0047$\\
~~~~$T_{FWHM}$\dotfill &FWHM transit duration (days)\dotfill &$0.0596^{+0.0021}_{-0.002}$&$0.1443^{+0.0048}_{-0.0047}$\\
~~~~$b$\dotfill &Transit Impact parameter \dotfill &$0.29^{+0.27}_{-0.20}$&$0.51^{+0.18}_{-0.27}$\\
~~~~$\rho_P$\dotfill &Density (cgs)\dotfill &$9.1^{+1.9}_{-2.6}$&$1.51^{+1.1}_{-0.86}$\\
~~~~$logg_P$\dotfill &Surface gravity \dotfill &$3.451^{+0.077}_{-0.13}$&$2.87^{+0.22}_{-0.36}$\\
~~~~$\Theta$\dotfill &Safronov Number \dotfill &$0.00474^{+0.00057}_{-0.00055}$&$0.0172^{+0.0096}_{-0.0089}$\\
~~~~$\fave$\dotfill &Incident Flux (\fluxcgs)\dotfill &$5.94^{+1.6}_{-0.7}$&$0.086^{+0.023}_{-0.01}$\\
~~~~$T_S$\dotfill &Time of eclipse (\bjdtdb)\dotfill &$2457394.2962\pm0.0013$ &$2457412.18^{+0.75}_{-1.2}$\\
~~~~$T_A$\dotfill &Time of Ascending Node (\bjdtdb)\dotfill &$2457394.4391\pm0.0013$&$2457402.44^{+0.62}_{-0.64}$\\
~~~~$T_D$\dotfill &Time of Descending Node (\bjdtdb)\dotfill &$2457394.1534\pm0.0013$&$2457408.77^{+0.48}_{-0.72}$\\
~~~~$M_P\sin i$\dotfill &Minimum mass ($M_{\oplus}$)\dotfill &$8.624^{+1.72}_{-2.00}$&$6.04^{+3.82}_{-3.18}$\\
~~~~$M_P/M_*$\dotfill &Mass ratio \dotfill &$0.0000264^{+0.000005}_{-0.0000036}$&$0.0000189^{+0.000011}_{-0.0000098}$\\
~~~~$d/R_*$\dotfill &Separation at mid transit \dotfill &$2.95^{+0.16}_{-0.34}$&$22.4^{+3}_{-3.9}$\\
~~~~$P_T$\dotfill &A priori non-grazing transit prob \dotfill &$0.333^{+0.043}_{-0.018}$&$0.0435^{+0.0091}_{-0.0051}$\\
~~~~$P_{T,G}$\dotfill &A priori transit prob \dotfill &$0.344^{+0.045}_{-0.018}$&$0.0458^{+0.0096}_{-0.0054}$\\
\hline
\end{tabular}
\label{tab:exofast_planetary}
\end{table*}

\section{Methods} \label{sec:methods}

\subsection{EXOFASTv2 Global Fits}
\label{sec:star}

To constrain the K2-106 system parameters, we performed a global fit of the \k2 photometry and RV data with the publicly available, exoplanet-fitting software {\tt EXOFASTv2} \citep{Eastman:2013,Eastman:2017,eastman:2019}. {\tt EXOFASTv2} can fit multiple light curves and radial velocity data sets and can model single and multi-planet systems using a differential evolution Markov Chain Monte Carlo algorithm.

\begin{figure}[ht]
\vspace{.1in}
\centering
\includegraphics[width=1\linewidth]{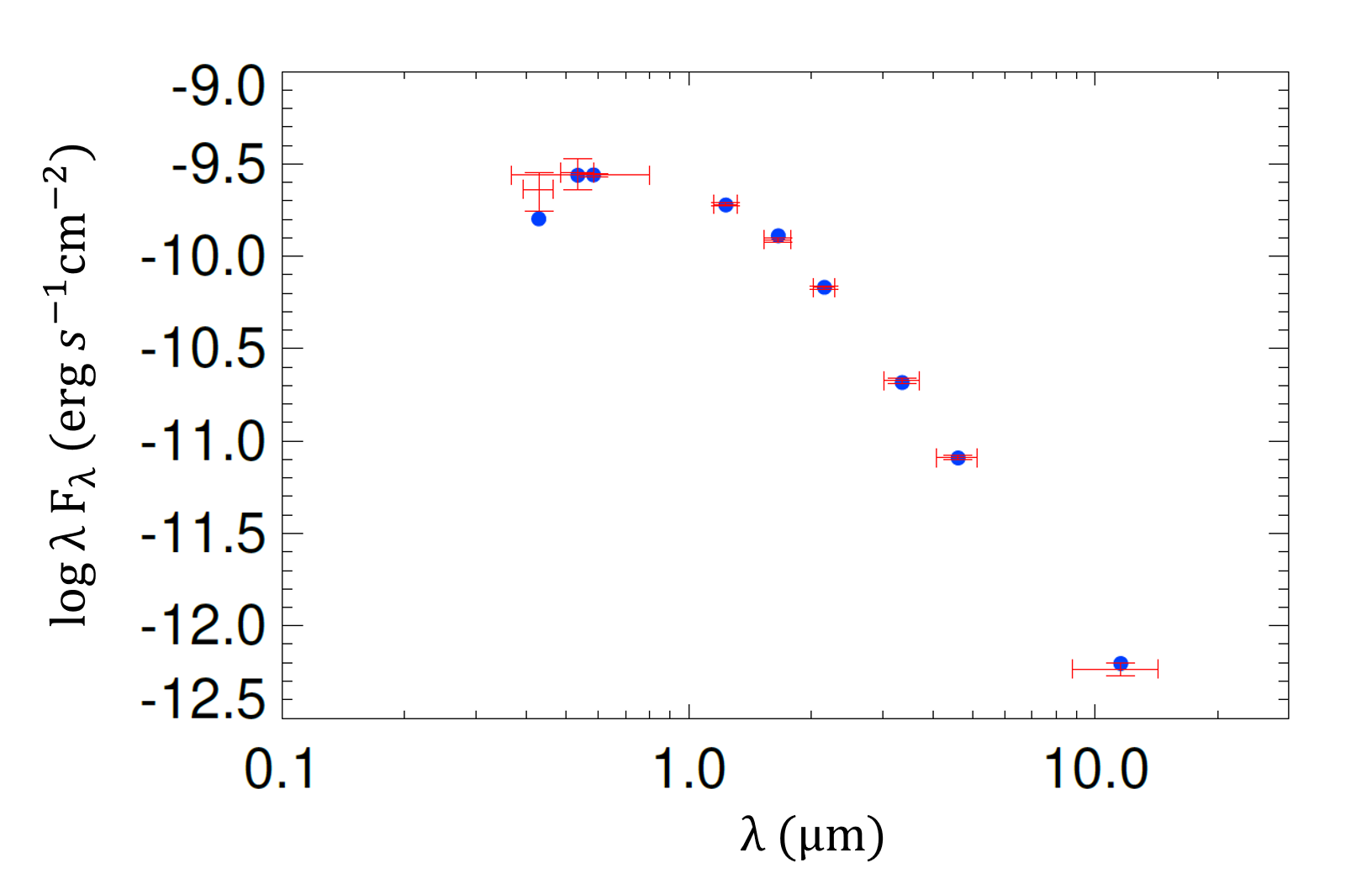}
\caption{Spectral Energy Distribution (SED) of K2-106. The red crosses show the broadband observations and the error bars show the width of the filters. The blue circles represent the model fluxes.}
\label{fig:sed} 
\end{figure}

We determined the host star properties by fitting  the star's spectral energy distribution (SED) simultaneously with the RV and photometric data using available broadband photometry; specifically the \textit{Gaia} $G$, $G_{BP}$, and $G_{RP}$, 2MASS $JHK$ magnitudes, and WISE W1-W4 magnitudes, these are all given in Table~\ref{tbl:LitProps}. We adopted a parallax of $\mu = 4.085 \pm 0.018$ mas from \textit{Gaia} EDR3. The fluxes were fit using the \citet{Kurucz:1992} atmosphere models. We set a prior on the maximum line-of-sight visual extinction $A_{V}$ of 0.17 from the \citet{Schlegel:1998} maps. We also adopted a wide [Fe/H] metallicity prior of $=0.06 \pm 0.2$ dex. This central value is the median of the metallicities reported in the literature for K2-106. The SED fit is shown in Figure~\ref{fig:sed}. For a detailed description of stellar SED fitting with EXOFASTv2, see, e.g., \citet{Godoy-Rivera}.

Using the resulting stellar properties from the SED fit as priors, we then performed a global fit of the system, using the \k2 photometry and the six RV data sets described in Section~\ref{sec:data}. We adopted an initial guess on the ephemererides of planets b and c from G17 of $T_{b} = 2457394.0114$ \bjdtdb~and $T_{c} = 2457405.7315$ \bjdtdb, and on their orbital periods of $P_{b} = 0.57$ days and $P_{c} = 13.33$ days, respectively. We assumed a zero eccentricity for K2-106b, which we justify as follows. The tidal circularization timescale, $\tau_{circ}$ for a planet of mass $M_p$, radius $R_p$, semi-major axis $a$, and tidal quality factor $Q$ is given by

\begin{equation}
    \tau_{circ}= \frac{4}{63}\frac{1}{\sqrt{GM_{\star}^3}}\frac{M_{p}a^{13/2}Q}{R_{p}^{5}}
\end{equation}

\citep{Goldreich:1966}. The tidal quality factor $Q$ for exoplanets is highly uncertain but we adopt a value of $Q=100$, which is reasonable for rocky planets (e.g., \citealt{Henning:2009}). Using the mass and radius from G17, we derive a tidal circularization timescale of $\tau_{circ} \leq 10,000$ years. Given the age that we infer for this system of $5.5^{+4.7}_{-3.7}$ Gyr, we deduce that enough time has passed for the planet's orbit to have been completely circularized, thus justifying our assumption that  $e=0$. We do not make the same assumption for K2-106c, however, given its longer orbital period of $\sim$13 days, but we impose a prior of $e = 0 \pm 0.2$.

We performed two global fits of the system: one using constraints from the MIST stellar evolutionary models and another without any constraints from stellar evolutionary models. The results of both analyses are shown in Tables~\ref{tab:mist_table} (\textit{with} constraints from MIST) and~\ref{tab:nomodel_table} (\textit{without}) and discussed in Section~\ref{sec:results}. The stellar and planetary parameters we derived for both the MIST and the model-independent fits are consistent with each other, as shown in Tables~\ref{tab:stellarprops} (stellar parameters) and~\ref{tab:planetaryprops} (planetary parameters).

\section{Results} \label{sec:results}

We first discuss the fit we performed without external constraints from stellar models. In order to break the well-known $M_{\star}-R_{\star}$ degeneracy in the fits of transiting planets \citep{Seager:2003}, we fit the star's SED using broad band photometry from publicly-available all-sky catalogs combined with the \textit{Gaia} EDR3 parallax, to provide a constraint on $R_{\star}$ (see Section~\ref{sec:star}). For this fit, we obtain a stellar mass of $M_{\star} =1.02^{+0.22}_{-0.32} M_{\odot}$, a radius of $ R_{\star} =0.99^{+0.029}_{-0.027} R_{\odot}$, a stellar effective temperature $T_{\rm eff} =  5489^{+75}_{-76}$ K and a density of $\rho_{\star} =1.49^{+0.26}_{-0.46}$ \dens. We derive a stellar mass and radial velocity semi-amplitude $K_{\star}$ consistent with G17, but derive a larger value for the stellar radius, leading to a planetary density of $\rho_{p} = 9.1^{+1.9}_{-2.6}$ \dens~(with a $\sim$25\% uncertainty), which is significantly lower than that derived by G17, who obtained $13.1^{+5.4}_{-3.6}$ \dens (with 34\% uncertainty). We conclude that, in agreement with the predictions from \citet{RodriguezMartinez:2021}, an analysis using purely empirical constraints on $R_{\star}$ and $M_{\star}$ results in a tighter constraint on the planetary density. We do not reach the predicted $\sim$16\% uncertainty on $\rho_{p}$, however, because our uncertainty on the ingress/egress time, $\tau$, is larger than the one found by G17 by a factor of $\sim$2.5, thus dominating the error budget on $\rho_{p}$ and leading to a slightly larger uncertainty than predicted in our previous paper. 

\begin{figure*}[ht]
\vspace{.1in}
\centering
\includegraphics[width=0.8\linewidth]{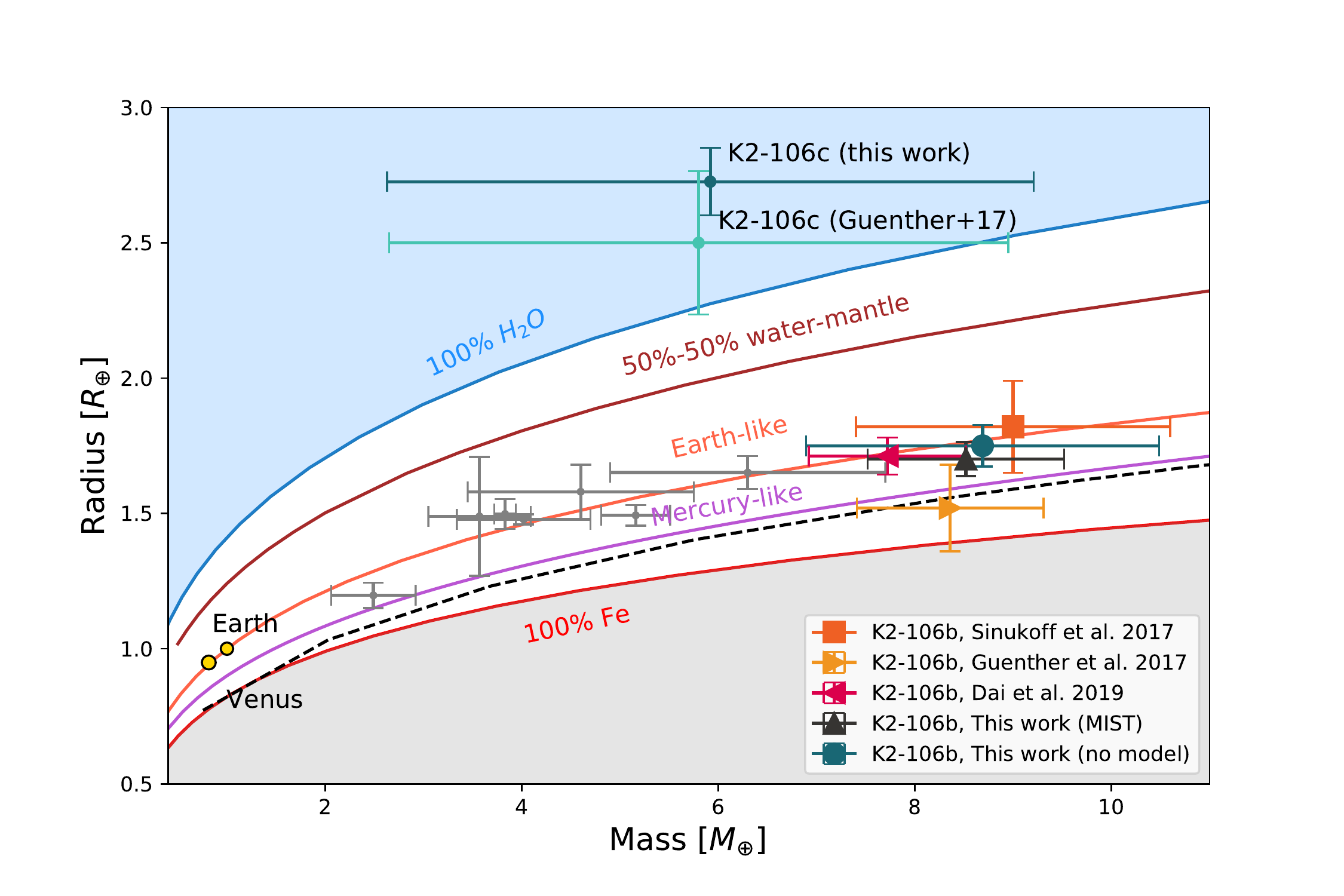}
\caption{Mass-radius diagram highlighting the measurements of K2-106b and K2-106c from this work and from previous literature. The colored lines are theoretical mass-radius curves from \citet{Zeng:2019}. Planets below the black dashed line would exceed the maximum iron allowed from collisional mantle stripping computed by \citet{Marcus:2010}. The gray points show the masses and radii of several other USPs for reference. Earth and Venus are also plotted for reference.}
\label{fig:mass-radius} 
\end{figure*}

We subsequently modeled the system additionally using priors on the host star from the MIST stellar evolutionary tracks and obtained consistent central values of the stellar and planetary parameters, but with slightly lower uncertainties than those derived from the fit without external constraints, as expected. From this fit, we obtain a planetary density of $\rho_{p}=9.4^{+1.6}_{-1.5}$ \dens, which is only slightly higher than, but consistent with, the one we derive without using MIST. We derive a stellar effective temperature of $T_{\rm eff} = 5508\pm70$ K and an isochronal age of $5.5^{+4.7}_{-3.7}$ Gyr. The temperature from both fits indicate that K2-106 is a late-type G star, either G7 or G8, based on the classifications of \citet{Pecaut:2013}. Both temperatures lead to a high equilibrium temperature for K2-106b of $T_{eq} = 2261^{+140}_{-70}$ K (from the model-independent fit), and $T_{eq} = 2275^{+36}_{-32}$ K (from the MIST fit). 

Our analysis of K2-106b indicates that although it is highly dense and possibly iron-enriched relative to Earth, it may not be a super-Mercury. Figure~\ref{fig:mass-radius} shows the mass and radius measurements of K2-106b and K2-106c, with other values from the literature plotted for reference. As can be seen from the plot, the measurement from G17 places K2-106b below the Mercury-like composition curve, implying an iron mass fraction possibly higher than Mercury's (of $\sim$70\%). The masses and radii from both \citet{Sinukoff:2017} and \citet{Dai:2019} imply a lower density and thus a more Earth-like composition. Finally, our measurements (with both MIST and the model-independent fit) place K2-106b decisively between the Earth-like and Mercury-like compositions. 

\begin{figure}[ht]
\vspace{.1in}
\centering
\includegraphics[width=1.1\linewidth]{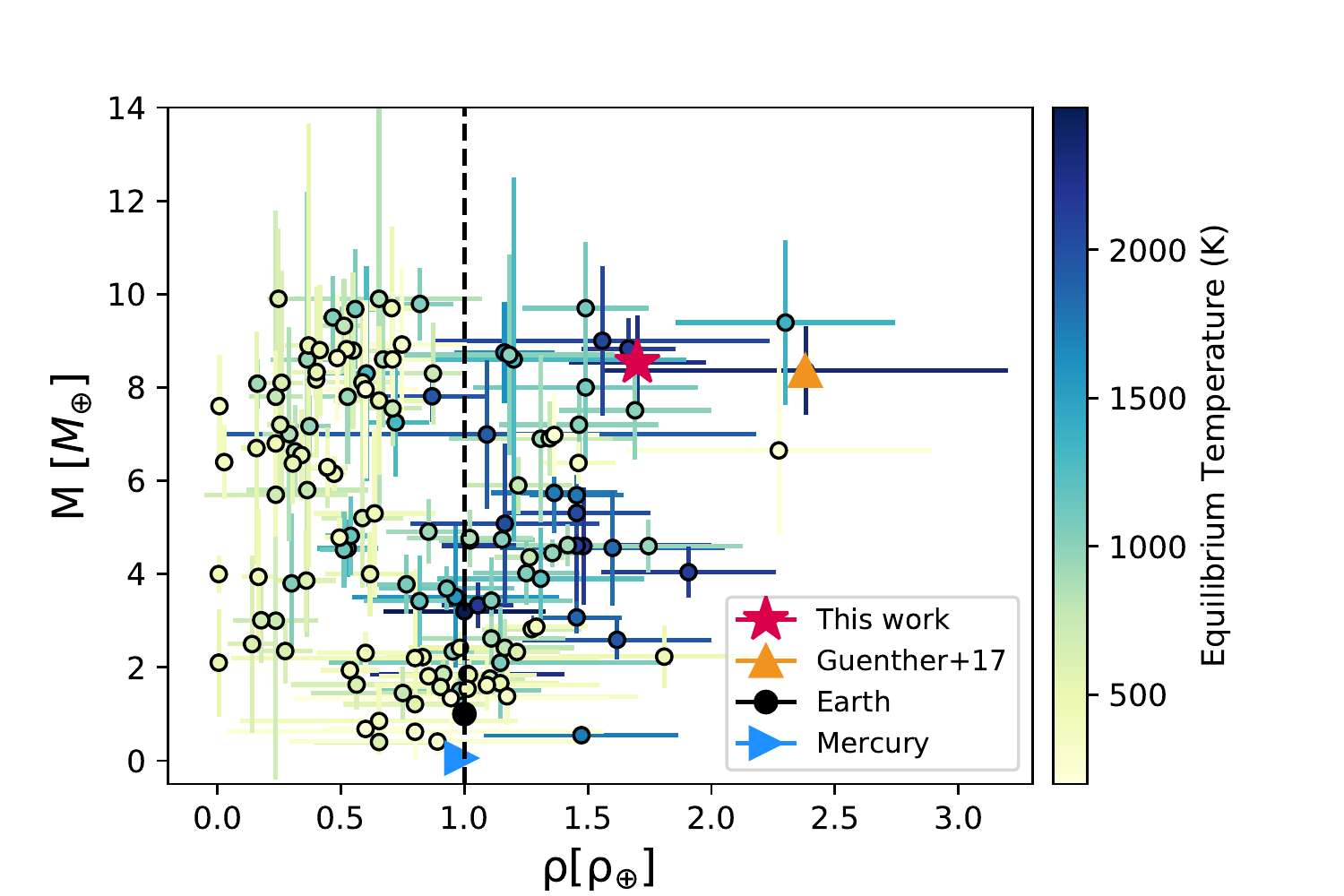}
\caption{Mass as a function of bulk density (normalized by the values of the Earth) for known planets with $M_{p}< 10~M_{\oplus}$. The dashed vertical line represents a constant Earth-like density. K2-106b is shown by a pink star, while the golden triangle is the value from \citet{Guenther:2017}. The points are color-coded by their equilibrium temperature. Earth and Mercury are over-plotted for reference (the black circle and blue triangle, respectively).}
\label{fig:density} 
\end{figure}

Figure~\ref{fig:density} shows the mass and density distribution of planets with $R_{p} <$ 2 $R_{\oplus}$ and $M_{p} < 10~M_{\oplus}$\footnote{https://exoplanetarchive.ipac.caltech.edu/}. These planets have an average bulk density of 6.59~\dens. With a density of $9.4^{+1.6}_{-1.5}$~\dens~(1.7$\rho_{\oplus}$), K2-106b ranks among the densest worlds known and is denser than 91\% of such planets. Also notable in Figure~\ref{fig:density} is a trend in density with equilibrium temperature: the denser, potentially rocky exoplanets ($\rho_{p}/ \rho_{\oplus} > 1$) are generally hotter than lighter, volatile-rich planets. This apparent dichotomy in bulk density as a function of equilibrium temperature (or semimajor axis) may support the theory that the densest planets form in the inner regions of the protoplanetary disk, where there is a greater supply of iron relative to other refractory elements (see, e.g., \citealt{Wurm:2013}).


\begin{table*}[ht]
\footnotesize
\setlength{\tabcolsep}{2pt}
\caption{Stellar Properties \label{tab:stellarprops}}
\centering
\begin{tabular}{lcccccc}
\hline\hline
Reference & $T_{\rm{eff}}$ & $\log{g_*}$ & $M_{\star}$ & $R_{\star}$ & $\rho_{\star}$   & [Fe/H] \\
& (K) & & ($M_{\odot}$) & ($R_{\odot}$) & (g/cm$^{3}$) & (dex)  \\
\hline
This work (no models) & 5489$^{+75}_{-76}$ & 4.46$^{+0.07}_{-0.16}$ & 1.02$^{+0.22}_{-0.32}$ & $0.99^{+0.029}_{-0.027}$ & 1.49$^{+0.26}_{-0.46}$ & $-$0.03 $\pm$ 0.01$^\dagger$  \\ 
This work (MIST) & 5508 $\pm$ 70 & $4.44^{+0.033}_{-0.037}$ & $0.96^{+0.063}_{-0.062}$ & $0.98^{+0.024}_{-0.022}$ & 1.46 $\pm$ 0.14 &  $-$0.03 $\pm$ 0.01$^\dagger$  \\ 
\citet{Adams:2017} & 5590 $\pm$ 51 & 4.56 $\pm$ 0.09 & 0.93 $\pm$ 0.01 & 0.83 $\pm$ 0.04 & -- & 0.02 $\pm$ 0.02 \\
\citet{Sinukoff:2017} & 5496 $\pm$ 46 & 4.42 $\pm$ 0.05 & 0.92 $\pm$ 0.03 & 0.95 $\pm$ 0.05 & -- & 0.06 $\pm$ 0.03 \\
\citet{Guenther:2017} & 5470 $\pm$ 30 & 4.53 $\pm$ 0.08 & 0.94 $\pm$ 0.06 & 0.87 $\pm$ 0.08 & -- & -0.02 $\pm$ 0.05 \\ 
\citet{Dai:2019} & 5496 $\pm$ 46 &	4.42 $\pm$ 0.05  & $0.90^{+0.057}_{-0.046}$ & $0.981^{+0.019}_{-0.018}$ & $1.35^{+0.14}_{-0.12}$ & 0.06 $\pm$ 0.03 \\ 

\hline

\end{tabular}

 \begin{flushleft} 
  \footnotesize{ 
    \textbf{\textsc{\hspace{0.3in}NOTES:}}
\citet{Adams:2017}, \citet{Sinukoff:2017}, and \citet{Guenther:2017} report $M_{\star}$ and $R_{\star}$, but do not report a stellar density, $\rho_{\star}$. We do not compute a stellar density from their masses and radii since the uncertainties in the latter are correlated. $^\dagger$This metallicity value was measured using a high-resolution spectrum (see Section~\ref{sec:abundances}), but we also infer a [Fe/H] abundance using the global fit with MIST constraints. These sets of abundances are consistent within their uncertainties but we adopt the more precise, spectroscopic [Fe/H].
               }
 \end{flushleft}

\end{table*}

\begin{table*}[ht]
\footnotesize
\setlength{\tabcolsep}{2pt}
\caption{Planet Properties \label{tab:planetaryprops}}
\centering
\begin{tabular}{lccccccc}
\hline\hline

Reference & Period & $K_{\star}$ & $R_{p}/R_{\star}$ & $R_{p}$ & $M_{p}$ & $\rho_{p}$ & \cmfrho  \\
& (days) & (m/s) & & ($R_{\oplus}$) & ($M_{\oplus}$) & (g/cm$^{3}$) & (\%) \\
\hline
This work (no models) & $0.571301^{+0.000015}_{-0.000016}$ &  $6.68 \pm 0.75$& $0.01617^{+0.0006}_{-0.00043}$ & $1.75^{+0.086}_{-0.068}$ & $8.69^{+1.69}_{-1.90}$ & $9.1^{+1.9}_{-2.6}$ & $39^{+19}_{-23}$ \\ 
This work (MIST) & $0.571302^{+0.000015}_{-0.000016}$ & $6.70^{+0.75}_{-0.76}$ & $0.01604^{+0.00043}_{-0.00040}$ & $1.71^{+0.069}_{-0.057}$ & $8.53\pm1.02$ & $9.4^{+1.6}_{-1.5}$ & $44^{+12}_{-15}$ \\ 
\citet{Adams:2017} & 0.571308 $\pm$ 0.000030 & --& $0.0161^{+0.0013}_{-0.0006}$ & 1.46 $\pm$ 0.14 & --& -- & -- \\
\citet{Sinukoff:2017} & 0.571336 $\pm$ 0.000020  &  7.2 $\pm$ 1.3 & $0.01745^{+0.00187}_{-0.00079}$ & $1.82^{+0.20}_{-0.14}$ & $8.57^{+4.64}_{-2.80}$ & -- & \\
\citet{Guenther:2017} & $0.571292^{+0.000012}_{-0.000013}$ & 6.67 $\pm$ 0.69  & $0.01601^{+0.00031}_{-0.00029}$ & 1.52 $\pm$ 0.16  & $8.36^{+0.96}_{-0.94}$ & $13.1^{+5.4}_{-3.6}$ & $80^{+20}_{-30}$ \\ 
\citet{Dai:2019} & 0.571 & $6.37^{+0.60}_{-0.62}$ & $0.01598^{+0.00056}_{-0.00057}$ &  1.71 $\pm$ 0.07  & $7.72^{+0.80}_{-0.79}$ &  8.5 $\pm$ 1.90 & 40 $\pm$ 23 \\ 

\hline
\end{tabular}

\end{table*}

\section{Composition of K2-106b} \label{sec:structure}

The relative abundances of the refractory elements (such as Fe, Mg, and Si) in the Sun have been shown to closely match those of Venus, Earth, and chondrite meteorites in the Solar System \citep{lodders:2009}. Many studies have since found strong chemical links between the relative abundances of the common, rock-forming elements Fe, Mg, and Si of the parent star and the compositions of their orbiting planets (see, e.g., \citealt{Dorn:2015, Brugger:2017, Plotnykov:2020,Schulze:2021,Liu:2020,Wilson:2021,Adibekyan:2021}). For example, \citet{Adibekyan:2021} found a positive correlation between the density of rocky planets and the iron mass fraction of their parent stars. Thus, the refractory elemental abundances of host stars can be used as a proxy for bulk composition and to constrain some properties of a planet's interior, such as the core/mantle mass and radius fractions. This is especially important because a planet's bulk density is degenerate with its interior composition.


\begin{figure*}[ht!]
\centering
\includegraphics[width=0.8\textwidth]{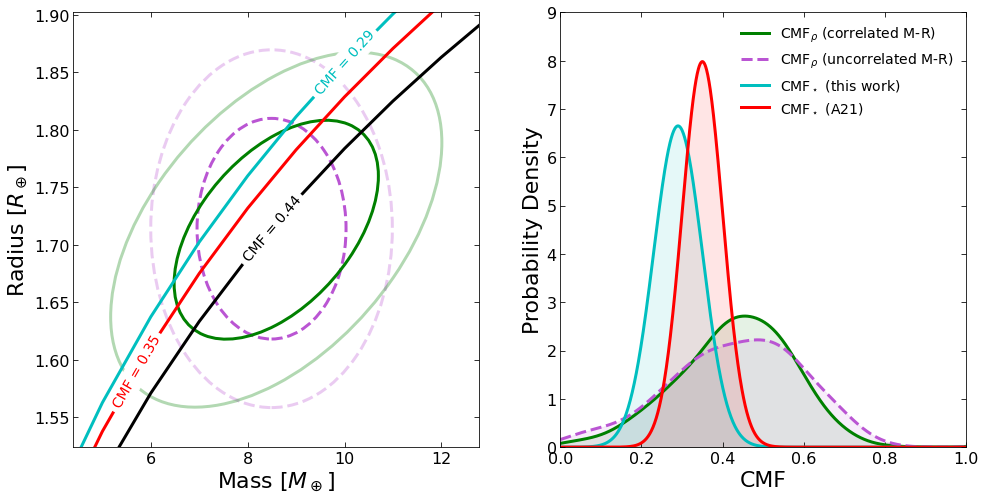}
\caption{\textbf{(Left)} $1\sigma$ and $2\sigma$ mass-radius ellipses for K2-106b. The violet ellipses assume that $M_{p}$ and $R_{p}$ are uncorrelated variables. The green ellipses represent the 1 and 2$\sigma$ errors when $M_{p}$ and $R_{p}$ are correlated via the planetary surface gravity, $g_{p}$. Planets along the blue, red, and black solid lines follow a \cmfrho~of 0.29, 0.35, and 0.44, respectively. The added constraint of $g_{p}$ reduces the uncertainty in the CMF. \textbf{(Right)} 1$\sigma$ distributions of the probability density functions (PDF) of the iron mass fraction or \cmfstar~for the abundances derived in this work (cyan) and for the \citet{Adibekyan:2021} values (red), and \cmfrho~outlined in purple (for uncorrelated $M_{p}$ and $R_{p}$) and green (for correlated).}
\label{fig:PDF}
\end{figure*}

One of the most important properties that characterizes the interior of a planet is its core mass fraction, defined as the mass of the core divided by the total mass of the planet. The CMF is a first-order approximation of its composition and a factor in the existence and lifetime of a magnetic field. The latter, in turn, is an essential ingredient for habitability, as it shields the planet from high-energy charged particles and reduces atmospheric loss. To constrain the CMF and interior structure of K2-106b, we combine the planet's physical properties with the star's atmospheric Fe/Mg and Si/Mg abundances. We follow the statistical framework of \citet{Schulze:2021}, which we briefly summarize here and direct the reader to for a more in-depth description.

\citet{Schulze:2021} calculate the CMF in two ways for a $\sim$dozen well-characterized, short-period planets with $R_p < 2R_{\oplus}$ and $M_p < 10 M_{\oplus}$. First, they calculate the CMF expected from the planet's mass and radius, denoted \cmfrho. Then they calculate the CMF as predicted from the refractory elemental composition of the photosphere of the host star, denoted \cmfstar~(and defined in Eqn~\ref{eqn:cmfstar}). They then calculate the probability that a planet satisfies the null hypothesis $H^{0}$, which says that the planet's composition reflects the chemical composition of its host star. This probability is denoted as $P(H^{0})$, and is quantified by the amount of overlap between the probability distributions of \cmfrho~and \cmfstar. In their framework, if P(\cmfrho $=$ \cmfstar) $\leq 32\%$, then the planet deviates from the expected composition at the $1\sigma$ significance level, and if P(\cmfrho $=$ \cmfstar) $\leq 5\%$, then it deviates from the composition expected from the host star at the $2\sigma$ significance level.

Here, \cmfstar~is mathematically defined as:

\begin{equation}
\label{eqn:cmfstar}
    \rm CMF_{\star} =  \frac{(\frac{Fe}{Mg})m_{Fe}}{(\frac{Fe}{Mg})m_{Fe} + (\frac{Si}{Mg})m_{SiO_{2}} + m_{MgO}}
\end{equation}

where X/Y are the stellar molar ratios for elements X and Y, and $m_i$ is the molar mass of species $i$. The uncertainty in \cmfstar~can be derived by propagating the uncertainties in Fe/Mg and Si/Mg; see Equation 3 of \citet{Schulze:2021}.

In the following subsections, we describe the chemical abundance determinations for K2-106 and the derivation of \cmfrho~and \cmfstar.

\subsection{Stellar Abundance Measurements}
\label{sec:abundances}

An estimate of \cmfstar~requires [Fe/H], [Si/H], and [Mg/H] abundances. We computed these using an archival optical HARPS spectrum spanning a wavelength range between 450$-$691 nm. We used the framework for spectral analysis, {\tt iSpec} \citep{blanco-cuaresma:14, blanco-cuaresma:19}, which calculates stellar atmospheric parameters as well as individual abundances using both the synthetic spectral fitting method and the Equivalent Width (EW) method. Within {\tt iSpec}, we used the radiative transfer code SPECTRUM \citep{Gray:1994}, the solar abundances from \citet{Grevesse:2007}, the MARCS.GES model atmospheres \citep{Gustafsson:2008}, and the atomic line list from the {\it Gaia}-ESO Survey (GES), which covers the entire wavelength of our spectrum. 

To constrain chemical abundances and their uncertainties via the spectral fitting method, we first calculated the spectrum's SNR with {\tt iSpec}, which yielded $\rm SNR\sim$37, with which we estimated the spectrum errors. We first derived the stellar atmospheric parameters with {\tt iSpec} using the synthetic spectral fitting method. We fixed the projected stellar rotation, \vsini, to 2 km/s and the limb darkening coefficient to 0.6 to avoid possible degeneracies between the macroturbulence and rotation, following the recommendations of \citet{blanco-cuaresma:14}. We derive the following parameters: $T_{\rm eff} = 5543.5 \pm 27.9$ K, $\log{g_*}= 4.41 \pm 0.05$, $\rm [M/H] = -0.01 \pm 0.02$ dex, $\alpha = 0.1 \pm 0.02$, microturbulent velocity $v_{\rm mic} = 1.54\pm0.05$ km/s, macroturbulent velocity $v_{\rm mac} =3.58 \pm 0.12$ km/s. To calculate the Fe, Mg, and Si abundances, we fixed all the stellar parameters to the values spectroscopically determined with {\tt iSpec} and only let the abundances vary. We opted to use the spectroscopic stellar parameters instead of the parameters derived from {\tt EXOFASTv2} because this ensures that both stellar parameters and abundances are calculated self-consistently. We note, however, that the stellar parameters derived with {\tt iSpec} and {\tt EXOFASTv2} are consistent within their uncertainties.

To account for systematics in the abundance determinations, we calculated the total uncertainty in each abundance as the quadrature sum of the uncertainties from the {\tt iSpec} errors and the contribution from atmospheric parameters $T_{\rm eff}$, $\log{g_*}$, and [M/H]. To estimate the systematic uncertainty from the atmospheric parameters, we performed a series of runs in which we fixed each stellar parameter ($T_{\rm eff}$, $\log{g_*}$, and [M/H]) to its $-2\sigma$, $-1\sigma$, $1\sigma$, and  $2\sigma$ values and calculated the [X/H] abundance at each value while fixing the other stellar parameters to their central values. Then, the systematic error in abundance from each stellar parameter was calculated as the deviation in the $2\sigma$ abundances from their central values, divided by 2. Our final abundances are $\rm [Fe/H] = -0.03 \pm 0.01$, $\rm [Mg/H] = 0.04 \pm 0.02$,  $\rm [Si/H] = 0.03 \pm 0.06$. To validate our measurements, we also calculated Fe, Mg and Si abundances with the code MOOG \citep{Sneden:2012}, which uses the equivalent width method, and obtained consistent sets of abundances. While writing this paper, we noticed that another group, \citet{Adibekyan:2021}, derived abundances of K2-106. We considered their results for the analysis of the composition of K2-106b, but we ultimately adopt the abundances from this work.  \citet{Adibekyan:2021} used a high-resolution HARPS spectrum of K2-106 and the code MOOG to determine stellar abundances and obtained consistent [Mg/H] and [Si/H] abundances but slightly higher [Fe/H] than we do (see Table~\ref{table:abundances}).

\begin{table*}[ht]
\footnotesize
\setlength{\tabcolsep}{2pt}
\caption{Abundance Measurements \label{table:abundances}}
\centering
\begin{tabular}{lcccccc}
\hline\hline
Source & [Fe/H] & [Mg/H] & [Si/H] & Si/Mg & Fe/Mg & \cmfstar \\
This work ({\tt iSpec}) & $-0.03 \pm 0.01$ & $0.04 \pm 0.02$ & $0.03 \pm 0.06$ & $0.93\pm 0.25$& $0.71\pm0.17$ & $0.29\pm0.06$ \\ 
This work (MOOG) & $-0.03 \pm 0.01$ & $0.03 \pm 0.02$ & $0.03 \pm 0.04$ & $0.95 \pm 0.24$  & $0.72 \pm 0.18$ & $0.29 \pm 0.06$  \\ 
\citet{Adibekyan:2021} & $0.10 \pm 0.03$ & $0.07 \pm 0.05$ & $0.05 \pm 0.03$ & $0.78\pm0.14$ & $0.85\pm 0.16$ & $0.35\pm0.05$ \\ 
\hline
\end{tabular}
\end{table*}

\subsection{Determination of \cmfrho~and \cmfstar}
\label{sub:cmfs}

We calculated the stellar molar ratios Fe/Mg and Si/Mg using Equation 5 from \citet{Hinkel:2018} and the associated uncertainties by propagating the errors from the abundance measurements. We calculate these ratios for three sets of abundance measurements: 1) the abundances derived in this work using {\tt iSpec}, 2) the abundances from MOOG, and 3) the abundances from \citet{Adibekyan:2021}. To determine \cmfstar, we insert the molar ratios in Equation~\ref{eqn:cmfstar}. 
The abundances, molar ratios, and values of \cmfstar~are all listed in Table~\ref{table:abundances}. All three abundance measurements lead to consistent values of \cmfstar~within the uncertainties.

We calculated \cmfrho~using the {\tt ExoPlex} program \citep{Unterborn:2018, Schulze:2021}, which self-consistently solves the equations of planetary structure including the conservation of mass, hydrostatic equilibrium, and the equation of state. The calculation of \cmfrho~assumes a pure, liquid iron core and a magnesium silicate ($\rm MgSiO_{3}$) mantle free of iron. This calculation takes as inputs the planetary mass and radius and their uncertainties as priors, as well as an initial guess for the \cmfrho~of the planet. We calculate two values of \cmfrho. Using the planetary parameters from our model-independent fit, we get \cmfrho $ = 39^{+19}_{-23}\%$. Using the parameters from the fit with MIST, we get a slightly higher value of \cmfrho $= 44^{+12}_{-15}\%$ as expected, since the planet density inferred from MIST is higher.

To complement the \cmfrho~values from {\tt ExoPlex}, we calculate the core-radius fraction (CRF) for K2-106b using the python package {\tt HARDCORE} \citep{Suissa:2018}, which uses boundary conditions to bracket the minimum and maximum CRF for a rocky planet, assuming a fully differentiated planet and a solid iron core. Once the CRF is known, the CMF can be approximated by $\rm CMF \approx CRF^{2}$ \citep{Zeng:2017}. For the planetary parameters from the empirical fit, we obtain a CRF of $64\pm17\%$, which corresponds to a CMF of 40 $\pm$ 21\%. Using the planetary parameters from the fit with MIST, we obtain a CRF of $68\pm 11\%$, corresponding to a CMF of 46 $\pm$ 15\%, both of which are fully consistent with the values from {\tt ExoPlex}.

\subsection{Is K2-106b Mercurified?} \label{sec:mercurified}

We find a value of $P(H^{0}) = 56.6\%$ for the \cmfrho~and \cmfstar~calculated in this work, and $P(H^{0})= 76.1\%$ for the \cmfstar~and \cmfrho~from the planetary mass and radius and stellar abundances from \citet{Adibekyan:2021}. Figure~\ref{fig:PDF}a shows the effect of correlating the planet's mass and radius via the surface gravity. As demonstrated in \citet{RodriguezMartinez:2021}, the constraint on the planet's surface gravity improves the precision on the planet's CMF, as shown by the green ellipses. Figure~\ref{fig:PDF}b shows the probability distribution functions and overlap between \cmfrho~and \cmfstar. Both stellar abundances and planetary parameters from this work or from \citet{Adibekyan:2021} lead to the conclusion that, given the uncertainties in the planet mass, radius, and host star abundances, there is insufficient evidence to robustly conclude that K2-106b is a true super-Mercury, under our formalism. We conclude that  lower stellar radius and thus lower planetary radius derived by previous authors led this planet to be misidentified as a super-Mercury in the literature. This highlights the importance of including the improved constraint on $R_{\star}$, which is due to the more precise Gaia DR3 parallax.

\subsection{Composition models of Acu\~na et al. 2021}
\label{sec:Acuna}

We complement the compositional analysis described in the previous section with the models from \cite{Acuna21}, which follow an MCMC Bayesian algorithm \citep{Acuna21,Director17}. The planet's interior is modeled with three layers: an Fe core, a Si-dominated mantle, and a hydrosphere in steam and supercritical states \citep{Mousis20,Acuna21}. When a water-dominated atmosphere is present, we compute its emitted radiation and Bond albedo to determine radiative-convective equilibrium with a 1D k-correlated model \citep[][Acu\~na et al. in prep]{Pluriel19}. The composition is therefore parameterized by two variables: the CMF and the water mass fraction (WMF). In our MCMC Bayesian analysis, the observables are the planetary mass and radius, and optionally the Fe/Si molar ratio. We consider three scenarios: in scenario 1, we assume that K2-106b is a completely dry planet (WMF=0), while the CMF is the only free non-observable parameter, and the mass and the radius are the observables. Scenario 2 is similar to scenario 1, but now the WMF is no longer constant, and it is left as a free parameter. Finally, in scenario 3, we include and treat the Fe/Si ratio as an observable. Its mean value and uncertainties are calculated from the host stellar abundances (see Table~\ref{table:abundances}). The resulting value is Fe/Si = 0.672 $\pm$ 0.094. The output of our interior Bayesian analysis are the posterior distributions of the compositional parameters (CMF and WMF) and the observables (mass, radius, and Fe/Si), which are shown in Table \ref{tab:interior_table}.

In scenario 1, the retrieved CMF is compatible with the estimates based on our approach from Section~\ref{sub:cmfs}, of CMF = 44$^{+12}_{-15}\%$. The planet's mass and radius are also reproduced by the model within their uncertainties. The 1$\sigma$ confidence intervals obtained by both interior models overlap with the interval of 23 -- 35\% for \cmfstar. We also explore if the estimated CMF by our model assuming the presence of an atmosphere is compatible with that of the star.

In scenario 2, the mean value of the CMF is higher than the CMF in scenario 1, but both are compatible within the uncertainties. In this scenario, the observed planetary mass and radius are also well reproduced by the model. A slightly higher CMF in scenario 2 is necessary to reproduce the observed mass and radius since a more Fe-rich bulk is more dense, leaving more space for a low-density, steam atmosphere to expand with respect to the dry scenario. As can be seen in Fig. \ref{fig:k2-106_mrdiag}, the estimated CMF in scenario 2 is compatible within uncertainties with the CMF estimated in our scenario 1 and by our previous estimate, although it is not compatible with CMF$_{\star}$. This means that if K2-106b was to reflect the CMF estimated from the abundances of its host star, it would be very unlikely to have a thick steam atmosphere, or indeed any atmosphere with a mean molecular weight lower than 18. If an atmosphere exists, the CMF would have to be greater than 0.46, which is the lower limit of the 1$\sigma$ confidence interval in scenario 2, and the atmosphere would have to be very thin. The properties of such a thin atmosphere would be $P_{\rm surf}$ = 184.9 $\pm$ 120.8 bar, $z_{\rm atm}$ = 404 $\pm$ 82 km, $T_{\rm surf}$ = 4154 $\pm$ 326 K and $A_{B}$ = 0.210 $\pm$ 0.001 of surface pressure, atmospheric scale height, surface temperature and Bond albedo, respectively.

In scenario 3, the MCMC Bayesian analysis can reproduce the Fe/Si molar ratio derived from the host stellar abundances, but the mass and radius are not compatible with their observed mean values and uncertainties. The retrieved CMF in scenario 3 is compatible with CMF$_{\star}$, as expected, since both estimates are calculated from the chemical abundances of the host star. In scenario 3, the retrieved mass is lower than the observed value, whereas the retrieved radius is higher than the observed one, yielding a lower planetary density (see Fig. \ref{fig:k2-106_mrdiag}, black dots in comparison with the yellow star and its uncertainties). This supports the conclusion we reached in scenario 2: K2-106b cannot have a steam atmosphere and reflect the refractory abundances of its host star simultaneously. Since any atmosphere with a lower mean molecular weight $\mu$ would have a higher scale height, we further conclude that K2-106b cannot have a substantial atmosphere with $\mu \lesssim 18$, which includes H and He-dominated atmospheres.

\begin{table}[h]
\centering
\begin{tabular}{lccc}
\hline
Parameter  & Scenario 1    & Scenario 2   & Scenario 3  \\ \hline
CMF        & 0.49$^{+0.16}_{-0.22}$  & 0.63$^{+0.19}_{-0.17}$  & 0.24 $\pm$ 0.03 \\
WMF & 0  & (5.8$^{+8.6}_{-5.8}$) $10^{-5}$ & (9.8 $\pm$ 7.8) $10^{-5}$ \\
$M_{p}$ [$M_{\oplus}$]   & 8.39$^{+1.05}_{-0.97}$ & 8.11$^{+1.36}_{-0.82}$ & 7.08$^{+0.59}_{-0.87}$  \\
$R_{p}$ [$R_{\oplus}$]   & 1.710$^{+0.069}_{-0.056}$ & 1.712$^{+0.068}_{-0.054}$ & 1.828$^{+0.043}_{-0.038}$ \\
Fe/Si      & 2.05 $\pm$ 1.52 & 4.75 $\pm$ 5.27 & 0.71 $\pm$ 0.11   \\ \hline
\end{tabular}
\caption{MCMC parameters of the interior structure analysis and their 1$\sigma$ confidence intervals for all three scenarios. In scenarios 1 and 2, our Bayesian model reproduces a CMF, $\rm M_{p}$, and $\rm R_{p}$ that is consistent with the results from our analysis in Section~\ref{sub:cmfs}. However, the model's retrieved CMF, mass and radius in scenario 3 do not reproduce well the observed values.}  
\label{tab:interior_table}
\end{table}

\subsection{Likelihood of a substantial atmosphere on K2-106 b and c} 
\label{sec:atmosphere}

In this section, we assess the probability that K2-106 b and c have extended atmospheres from a different angle. Given both the high density ($9.4^{+1.6}_{-1.5}$ \dens) and orbital period ($P= 0.57$ days) of K2-106b, it is probable that any primordial atmosphere it may have accreted during formation has been lost to atmospheric escape. We consider this possibility by estimating the XUV-driven mass loss rate of the atmosphere over the planet's lifetime.  First, we consider the restricted Jeans escape parameter $\Lambda$  \citep{Fossati:2017}, defined as
\begin{equation}
    \Lambda= \frac{GM_{p}m_{H}}{k_{B}T_{eq}R_{p}}
\end{equation}

where $m_{H}$ is the mass of a hydrogen atom, $M_p$, $R_p$ and $T_{\rm eq}$ are the mass, radius and equilibrium temperature of the planet, and $k_{B}$ is Boltzmann's constant. We calculate a restricted Jean's escape value of 16.5 for K2-106b. \citet{Cubillos:2017} found that planets with restricted Jean's escape parameters $\Lambda \leq 20$  cannot retain hydrogen-dominated atmospheres and are in the  ``boil-off" regime. Below this value,  atmospheric escape is mainly driven by a combination of planetary thermal energy and low gravity rather than by high irradiation \citep{Sanz-Forcada:2011}. 

Next, we estimate the mass loss rate, $\dot{M}$ --assuming that K2-106b formed with a substantial atmosphere and that the main driver of atmospheric escape is a combination of planetary intrinsic thermal energy and low gravity-- using the following equation from \citet{Kubyshkina:2018}: 

\begin{equation}
    \frac{\dot M}{\rm g~s^{-1}} = e^{\beta}(F_{XUV})^{\alpha_{1}}\bigg(
\frac{d}{AU}\bigg)^{\alpha_{2}}\bigg(
\frac{R_{p}}{R_{\oplus}}\bigg)^{\alpha_{3}} \Lambda^{K}
\end{equation}

where the coefficients in the equation are determined by whether $\Lambda$ is above or below a certain value and are given in \citet{Kubyshkina:2018}. We estimate the XUV flux, $F_{\rm XUV}$, using the age-luminosity relations from \citet{Sanz-Forcada:2011}. For K2-106b, we expect $\dot{M} = 1.28\times 10^{11}$ g/s. Assuming that the initial mass of the atmosphere is 1\% of the total mass of the planet, the atmosphere would be completely evaporated in $\sim$126 Myr. Given the current age of the system of $\sim$5.5 Gyr, K2-106b should have lost any primary atmosphere it may have had. This result agrees with the findings of \citet{Fossati:2017}, who predicted that planets with  masses lower than $5M_{\oplus}$, equilibrium temperatures higher than 1000 K, and $\Lambda$ between 20$-$40, would have atmospheres that would evaporate in $\lesssim 500$ Myr. Therefore, in agreement with \citet{Fossati:2017} and G17, we posit that K2-106b is unlikely to have a substantial atmosphere.

For K2-106c, we derive a density of $1.58^{+0.96}_{-0.84}$ \dens~for the global fit with constraints from MIST and $1.51^{+1.1}_{-0.86}$ \dens~for the model-independent fit. The relatively low mass and bulk density of K2-106c suggests that it could be a low-density super-Earth with an extended hydrogen atmosphere, or contains a substantial fraction of water (an ``ocean world", \citealt{kuchner:2003,Leger:2004}). However, the large uncertainty in the planet's mass ($\Delta m_p/m_p \sim 56\%$) makes it difficult to constrain its structure, and it is possible that the mass in this work and the literature have been underestimated, leading to the low bulk density measured. G17 estimated a mass loss rate $\dot{M}= 4\times10^{9}$ g/s and $\Lambda = 28.8 \pm 9.2$ and concluded that the planet may be in the ``boil off" regime, and, at its estimated atmospheric escape rate, K2-106c should have lost its atmosphere in a few Myr, similar to its hotter sibling. Given the uncertainty in its mass, more observations will be needed to resolve the composition of K2-106c.

These conclusions may be tested with future follow-up atmospheric observations. Given its short period, K2-106b's orbit is likely to have circularized (as we show in Section~\ref{sec:star}) and also to be tidally locked to its host star. In the absence of an atmosphere, tidal locking would lead to high temperature contrasts between the dayside and the nightside, with differences between both sides of the order of 1000 K. Nevertheless, if K2-106b possesses an atmosphere, winds from the hot dayside could distribute the heat to the cool nightside and thus partially homogenize the planet's surface temperature. One way to confirm or rule out the presence of an atmosphere could be to obtain a phase curve of the planet with JWST. Recently, \citet{Kreidberg:2019} obtained phase curves with Spitzer of the low-mass ultra-short period planet LHS-3844b and derived a dayside temperature of 1040 $\pm$ 40 K and a temperature consistent with zero Kelvin on the nightside, ruling out the presence of an atmosphere on the planet. They reasoned that if LHS-3844b possessed a substantial atmosphere, it could distribute the heat from the dayside to the nightside and thus we would not observe such a large day- and night-side temperature variation. They also modeled the emission spectra of several rocky surfaces and compared them to the planet-to-star flux they measure for the planet, and conclude that its reflectivity is most consistent with a basaltic composition, i.e., with a crust formed from volcanic eruptions. This is a possible indication that the planet is a so-called ``lava planet", a hypothesized world characterized by high dayside equilibrium temperatures ($T_{\rm eq}$ between 2500$-$3000 K) and surfaces covered by molten lava. Given the density, period, and the fact that it's likely tidally locked, K2-106b may well be a lava planet, which could be investigated by replicating the analysis of \citet{Kreidberg:2019} to indirectly infer the presence or lack of an atmosphere (see also \citealt{Keles:2022} and \citealt{Zieba:2022}). Additionally, if K2-106b is indeed a lava world, then perhaps magma oceans could have led to the formation of a secondary atmosphere of heavy elements through outgassing of volatiles in the interior (see, e.g., \citealt{Papuc:2008}), or a mineral atmosphere comprised of Na, Mg, O, or Si, (see e.g., \citealt{Ito:2021}). Whether K2-106b has an escaping atmosphere, acquired one via outgassing through tectonic activity, or does not have an atmosphere at all can be further investigated with follow-up observations and thus the K2-106 system provides an excellent laboratory for atmospheric studies.

\section{Conclusions} \label{sec:conclusions}

In this paper, we revisited the K2-106 system and improved upon the physical properties of its two known transiting exoplanets. We employed planet interior models to demonstrate that, even though K2-106b is iron-enriched based on its high density and relatively large core mass fraction, it is indistinguishable from its host stars' Fe/Mg ratios given observational constraints, which implies that it did not suffer the formation processes responsible for Mercury's large core. Given its mass, radius, and chemical abundances of the parent star, K2-106b is unlikely to be a super-Mercury, as was previously thought. In addition, its high bulk density and proximity to its host star implies that K2-106b probably lacks a primordial or even secondary atmosphere and could perhaps be a barren lava world. In contrast, the low density of the outer planet, K2-106c ($\rho_{p} = 1.58^{+0.96}_{-0.84}$~\dens), suggests that it could be a water world or have an extended H/He atmosphere. These hypotheses could potentially be tested with future follow-up observations with JWST. 

To robustly characterize the interior structure and composition of low-mass planets, we will need extremely precise masses and radii with uncertainties on the order of $<10\%$ and $\sim2\%$. Reaching these precisions is often difficult even with extensive transit and radial velocity observations, as evidenced by this work. In addition, planet mass and radius alone are not sufficient to determine the interior structure of a planet, as is well known, and thus the refractory elemental abundances of the host star are extremely useful as a proxy for planet composition. Based solely on its bulk density, one might incorrectly deduce that K2-106b is a super-Mercury. Yet, upon more careful analysis, and after considering the refractory elemental abundances of the host star, it is apparent that the composition of K2-106b is more Earth-like and is consistent with the photospheric abundances of its host star. One corollary of this study, then, is that perhaps other putative super-Mercuries in the literature may not be as iron-enhanced as they seem. The characterization of super Mercuries, as well as highly irradiated, ultra short period planets, is critical to constrain the frequency and compositional diversity of small planets, and we therefore encourage more observations of K2-106 and similar systems.

\acknowledgments

RRM and BSG were supported by the Thomas Jefferson Chair for Space Exploration endowment from The Ohio State University. RRM thanks Joseph Rodriguez, Jason Eastman, and Tharindu Jayasinghe for valuable discussions throughout this project. J.W. acknowledges the support by the National Science Foundation under Grant No. 2143400. We acknowledge the academic gift from Two Sigma Investments, LP, which partially supports this research. This research has made use of the NASA Exoplanet Archive, which is operated by the California Institute of Technology, under contract with the National Aeronautics and Space Administration under the Exoplanet Exploration Program. This work has made use of data from the European Space Agency (ESA) mission {\it Gaia} (\url{https://www.cosmos.esa.int/gaia}), processed by the {\it Gaia} Data Processing and Analysis Consortium (DPAC,
\url{https://www.cosmos.esa.int/web/gaia/dpac/consortium}). Funding for the DPAC
has been provided by national institutions, in particular the institutions
participating in the {\it Gaia} Multilateral Agreement.

\software{{\tt EXOFASTv2} \citep{eastman:2019}, {\tt ExoPlex} \citep{Unterborn:2018},
          \wotan~ \citep{Hippke:2014},
          {\tt iSpec} \citep{blanco-cuaresma:14,blanco-cuaresma:19}, {\tt HARDCORE} \citep{Suissa:2018}, numpy \citep{numpy}.}

\clearpage

\bibliography{sample63}{}
\bibliographystyle{aasjournal}

\appendix

\begin{figure}[h]
\centering
\includegraphics[scale=0.8]{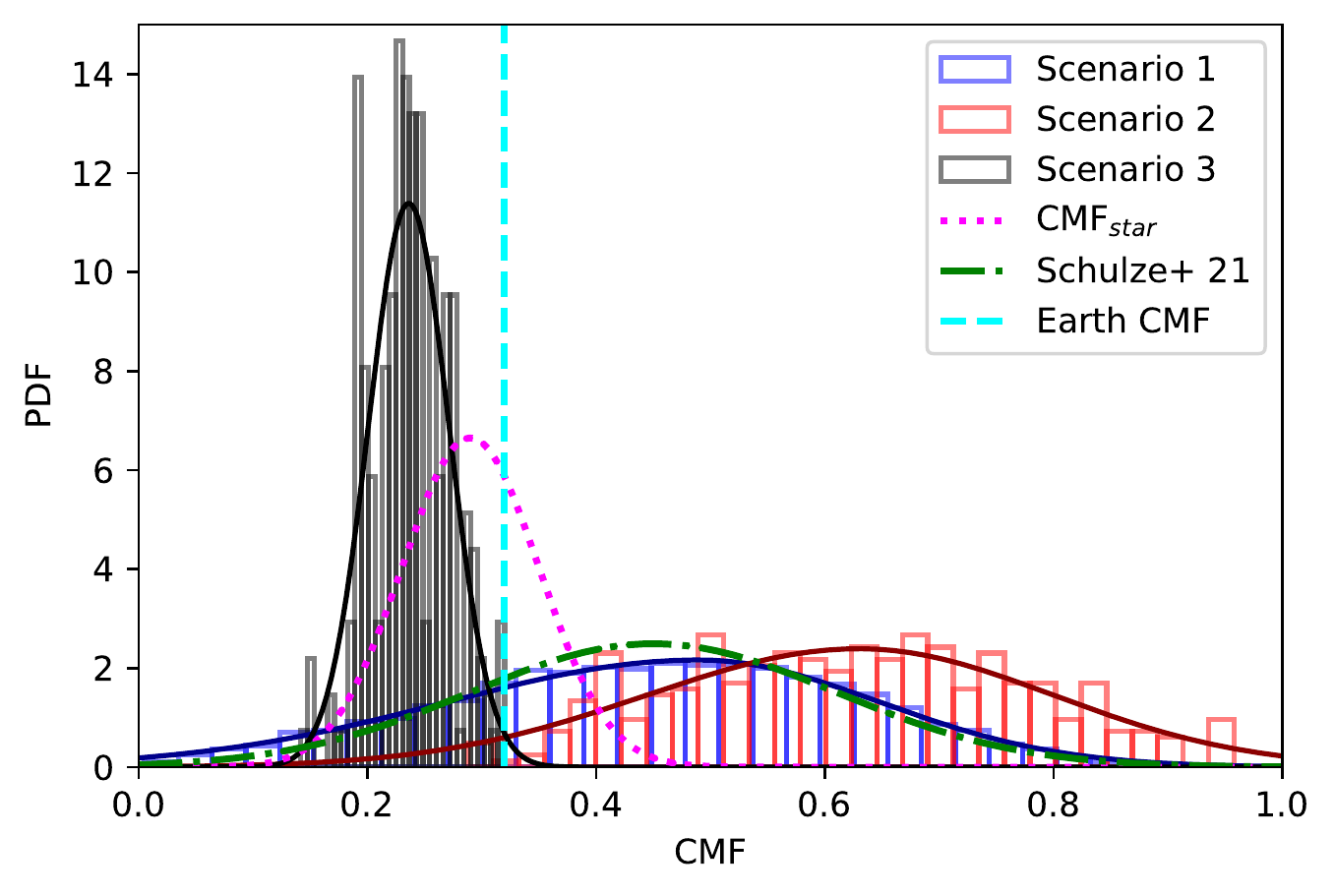}
\caption{CMF posterior distribution functions for the three scenarios presented in Section~\ref{sec:Acuna}. The distributions for \cmfstar~and the CMF estimated by the interior model of \cite{Schulze21} under the assumption of a dry planet, similar to our scenario 1, are shown for comparison.}
\label{fig:CMF_distr}
\end{figure}

\begin{figure}[h]
\centering
\includegraphics[scale=0.5]{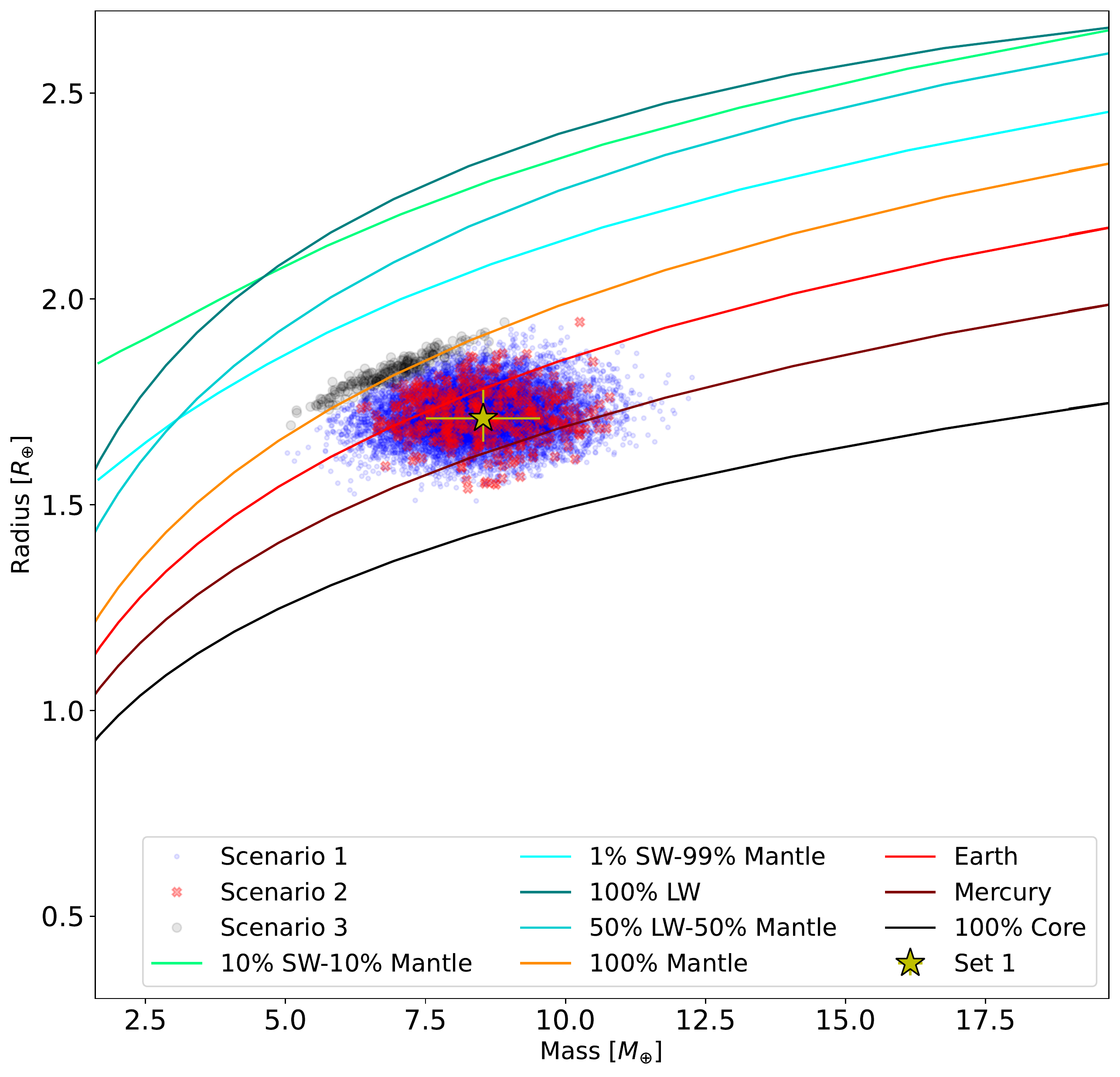}
\caption{Mass-radius diagram for K2-106b. The realizations of the MCMC for the three scenarios from Section~\ref{sec:Acuna} are shown, together with the mass-radius relationships for planets with varying CMFs and liquid WMFs (LW) from \cite{Brugger17}, and supercritical water (SW) from \cite{Acuna21}. For the latter, we assume an equilibrium temperature of 1200 K.}
\label{fig:k2-106_mrdiag}
\end{figure}

\end{CJK*}
\end{document}